\documentclass[preprint,english,showpacs,12pt,floatfix]{revtex4}
\usepackage{babel,amsmath,amssymb,dcolumn}
\usepackage{hyperref}
\usepackage{amsfonts}
\usepackage{amssymb}
\usepackage{graphicx}
\usepackage{latexsym}
\usepackage{dcolumn}
\usepackage{epsfig}
\usepackage{subfigure}
\begin{document}

\newcommand{\vp}{\varphi}
\newcommand{\nn}{\nonumber\\}
\newcommand{\beq}{\begin{equation}}
\newcommand{\eeq}{\end{equation}}
\newcommand{\bed}{\begin{displaymath}}
\newcommand{\eed}{\end{displaymath}}
\def\bea{\begin{eqnarray}}
\def\eea{\end{eqnarray}}
\newcommand{\veps}{\varepsilon}
\newcommand{\nablasl}{{\slash \negthinspace \negthinspace \negthinspace \negthinspace  \nabla}}
\newcommand{\om}{\omega}

\newcommand{\Dsl}{{\slash \negthinspace \negthinspace \negthinspace \negthinspace  D}}
\newcommand{\tDsl}{{\tilde \Dsl}}
\newcommand{\tnablasl}{{\tilde \nablasl}}
\title{Fermion perturbations in string-theory black holes II: the higher dimensional case}

\author{Owen Pavel Fern\'{a}ndez Piedra}
\email{opavel@ucf.edu.cu, opavelfp2006@gmail.com }
\affiliation{$^{1}$ Grupo de Estudios Avanzados, Universidad
de Cienfuegos, Carretera a Rodas, Cuatro Caminos, s/n. Cienfuegos,
Cuba,}
\affiliation{$^{2}$ Instituto de F\'{\i}sica, Universidade de S\~ao Paulo,
  CP 66318,
05315-970, S\~ao Paulo, Brazil}

\author{J. Bernal Castillo, Y. Jim\'enez Santana and L. Figueredo Noris }
\affiliation{$^{2}$ Grupo de Estudios Avanzados, Universidad
de Cienfuegos, Carretera a Rodas, Cuatro Caminos, s/n. Cienfuegos,
Cuba}

\begin{abstract}
In this paper we report the results of a detailed investigation of the complete time evolution of massless fermion fields propagating in spacetimes of higher dimensional stringy black hole solutions, obtained from intersecting branes in string/$M$ theory. We write the Dirac equation in $D$-dimensional spacetime in a form suitable to perform a numerical integration of it, and using a Prony fitting of the time domain data, we determine the quasinormal frequencies that characterize the test field evolution at intermediary times. We also present the results obtained for the quasinormal frequencies using a sixth order WKB approximation, that are in perfect agreement with the numerical results. The power law exponents that describe the field relaxation at very late times are also determined, and we show that they depends upon the dimensionality of space-time, and are identical to that associated with the relaxation of boson fields for odd dimensions. The dependance of the frequencies and damping factor of the spinor field with the charges of the stringy black hole are studied.
\end{abstract}

\preprint{GEA-UCF 2012-02}
\date{\today}
\maketitle

\section{Introduction}

The study of perturbations around black hole solutions has a long history, beginning with the work of Regge and Wheeler in 1957 \cite{Regge:1957td}. They addressed the problem of finding the effects that the evolution of small metric disturbances can produce on the stability of Schwarzschild black holes. To solve the problem, they performed a split of the metric perturbation according to the behavior of the different components under the symmetry group of the background space-time itself, leading to a classification of disturbances in different sectors. The same idea have been used since that time, and a variety of perturbation problems has been addressed for a number of interesting solutions, coming from Einstein General Theory of Relativity to String Theory \cite{Kokkotasreview,Nollertreview,bertireview,zhidenkoreview}.

Apart from the obviously important question about the stability of a gravitational system, the study of perturbations of black hole solution can serve to know about the geometry of the background space-time. It is known that the time evolution of any matter field perturbation comes in three stages: an initial transient stage in which the black hole begins to absorb the disturbance, and that is highly dependent on the initial form of the fluctuation, followed by a set of exponentially damped oscillations called quasinormal modes, and, at very large times, late time tails depending of the asymptotic form of the effective potential describing the perturbation.

The above picture is typical for massless perturbations in asymptotically flat space-times, where the late time tails have a power law character. For black holes in de Sitter backgrounds, the tail depends on the magnitude of the cosmological constant $\Lambda$. For large values of the cosmological constant the late time evolution of disturbances is characterized by exponential tails, whereas for small $\Lambda$ exponential and power law tails can appear. For negatives values of cosmological constant, we have usually a time evolution completely dominated by quasinormal modes, but in some asymptotically Anti de Sitter (AdS) space-times describing charged black holes, there can appear an exponential falloff \cite{zhidenkothesis,tesismolina,mitesis}.

The quasinormal modes carry information of the black hole space-time geometry, and can serve to estimate parameters as the mass, charge and angular momentum of black holes \cite{berti3}.

Quasinormal modes are also important in such contexts as String/M theory applications, in which the celebrated AdS/CFT correspondence can allow us to study the properties of strongly interacting systems \cite{maldacenaadscft,aharony,Gubser:2009md}. Related to this general idea, a more specific setup called gauge/gravity duality establish that a black hole in asymptotically AdS space-time is dual to a strongly interacting system defined at the AdS boundary, and the imaginary part of the fundamental quasinormal mode is proportional to the thermalization time in the boundary theory \cite{horowitz-hubeny,kovtun-starinets}. Gauge/ gravity correspondence have been applied in to study a great variety of physical systems, from quark-gluon plasmas to holographic superconductors \cite{Son:2007vk,Hartnoll:2009sz}. Such idea can be extended to systems without Lorentz invariance, as the recently discovered asymptotically Lifshitz black holes, giving us the possibility to study strongly correlated systems coming from condensed matter theory \cite{eloyhdim,kachru1}.

This is the second of a series of papers devoted to the study of
perturbations of a family of black holes coming
from string theory \cite{cvetic1,cvetic2,cvetic3}. In this work we extend the results of our previous study of four dimensional stringy black holes \cite{owenjeferson3} to higher dimensional cases. We continue considering the evolution of fermion perturbations in the form of massless uncharged
Dirac fields.

The motivations to this work coming from the fact that higher dimensional black hole solutions appears as natural ones when we consider a String/M theory setup, and for this reason it is important to know if the solutions are stable against
small perturbations in order to make they more reliable. Thus, the propagations of disturbances of different spin weight is an important subject to study in such backgrounds. As fermions universally describe some matter fields, in most cases they are essential
for the structure of the solutions, raising the importance to study fermion fluctuations \cite{gibbons-rogatko,splitfermion}.

Also the analysis of the quasinormal spectrum of the higher dimensional black holes in
string/M theory provides a simple way to fix the parameters of the solutions, and consequently, of string/M Theory itself. Another interesting point is related with the ideas of AdS/CFT and holography. In this context, fermion perturbations of the gravitational bulk fields equations, turn out
to be fundamental matter fields in the boundary conformal field
theory, providing an important class of objects relevant for
condensed matter physics.

The paper is organized as follows: Section II briefly presents the line element of higher dimensional black hole obtained from
intersections of branes. In Section III we obtain a general wave equation
suitable to analyze the propagation of a  massless Dirac field in the
background geometry of the higher dimensional black holes. In Section IV we applied the above theory to study numerically the evolution equation of
such fields in the considered backgrounds, and compute the complex
quasinormal frequencies using two methods: Prony fitting of time
domain data and the sixth order WKB approach. We also presents the results of a
numerical investigation of the relaxation of Dirac perturbations in higher dimensional
stringy black holes at very long times. In Section V we consider the large angular momentum limit, and  write an analytical formula for the quasinormal frequencies. Finally, after the concluding remarks, we include an Appendix with Tables showing some numerical results obtained in our study.

\section{D-dimensional non-extreme stringy black hole solutions}

A particular class of solutions that codify information from string/M-theory and that can be regarded as black holes in it, are those non-extremal solutions that appear as a result of the intersection of branes in M-theory \cite{cvetic1,cvetic2,cvetic3}. The parameters in those stringy black holes solutions can be deduced from the compactification of higher dimensions. The main idea of such compactified configurations is to relate static extremal/non-extremal solutions with extreme/non-extreme versions of solutions coming from brane surgery. This approach allow us to know about the structure of non-extremal black holes, shedding some light about the origin of entropy in such systems \cite{cvetic1}.

In D space-time dimensions, the line that describes the geometry of stringy black holes is \cite{cvetic1,cvetic2,cvetic3}
\begin{equation}\label{metric}
ds^{2}=h(r)^{-\frac{D-3}{D-2}}f(r)dt^{2}+h(r)^{\frac{1}{D-2}}f(r)^{-1}dr^{2}+r^{2}h(r)^{1/D-2}d\Omega_{2}^{D-2},
\end{equation}
where
\begin{equation}\label{f}
f(r)=1-\left(\frac{r_{H}}{r}\right)^{D-3}.
\end{equation}
and
\begin{equation}\label{h}
h(r)=\prod_{i=1}^{n_{D}}\left[1+\left(\frac{r_{H}}{r}\right)^{D-3}Q_{i}\right]
\end{equation}
The above metrics describes solutions parametrized by $r_{H}=\mu^{\frac{1}{D-3}}$, the position of the event horizon, and $n_{D}$ charges $Q_{i}=\sinh^{2}(\delta_{i})$, that can be written in terms of the parameters $\delta_{i}$, coming from the compactification of additional dimensions. The quantity  $n_{D}$ depends upon the space-time dimensions, being $n(4)=4$, $n(5)=3$ and $n(6\leq D \leq 9)=2$. In the above definition of $r_{H}$, $\mu$ is the so called non-extremality parameter, that also contain information from the higher compactified dimensions.

\section{Massless spinor perturbation in spherically symmetric space-times}

In this section we write the massless Dirac equation in a general $D$-dimensional curved space-time in a form suitable for its numerical and analytical study.
In the following we consider a \(D\)-dimensional manifold
\(\mathcal{M}\), whose line element can be written locally as a warped product of the form
\begin{equation}\label{warpedproduct1}
    \mathcal{M}^{D}=\mathcal{O}^{2}\times\mathcal{K}^{D-2},
\end{equation}
We introduce coordinates $z^{\mu}=(y^{a},x^{i})$ in $\mathcal{M}^{D}$, where $y^{a}$ and $x^{i}$ are coordinates defined in the manifolds $\mathcal{O}^{2}$ and $\mathcal{K}^{D-2}$, respectively. In terms of this coordinates, the full metric is given by
\begin{equation}\label{warpedmetric2}
    ds^{2}=\mathfrak{g}_{\mu\nu}dz^{\mu}dz^{\nu}=g_{ab}(y^{a})dy^{a}dy^{b}+w^{2}(y^{a})d\sigma_{D-2}^{2},
\end{equation}
where $w(y^{a})$ is an arbitrary function defined in $\mathcal{O}^{2}$ and
\begin{equation}\label{maxspacemetric1}
    d\sigma_{n}^{2}=\gamma_{ij}(x)dx^{i}dx^{j},
\end{equation}
is the metric on \(\mathcal{K}^{D-2}\), that we assume is an Einstein manifold. We also consider that $\mathcal{N}^{m}$ is lorentzian.
When the line element (\ref{warpedmetric2}) describes a black-hole space-time, then the space \(\mathcal{K}^{n}\) describes the structure of a spatial section of its event horizon. If \(\mathcal{K}^{n}\) is a constant curvature space with sectional curvature \(K\), then \(K=0\) represents a flat space, \(K=1\) a spherical manifold and  \(K=-1\) an hyperbolic one \cite{ik}.

In curved space-time the massless Dirac equation is
\begin{equation}\label{}
  \nablasl \Psi=0,
\end{equation}
where $\nablasl=\Gamma^{\mu}\nabla_{\mu}$ is the Dirac operator that
acts on the spinor $\Psi$, $\Gamma_{\mu}$ are the curved space
Gamma matrices, and the covariant derivative is defined as
$\nabla_{\mu}=\partial_{\mu}-\frac{1}{4}\om_{\mu}^{a
b}\gamma_{a}\gamma_{b}$, with $a$ and $\mu$ being tangent and
space-time indices respectively, related by the basis of orthonormal
one forms $\vec{e}^{\ a}\equiv e_{\mu}^{a}$. The associated
conection one-forms $\om_{\mu}^{a b}\equiv\om^{a b}$ obey
$d\vec{e}^{\ a}+\om^{a}_{\ b}\wedge\vec{e}^{\ b}=0$, and the
$\gamma^{a}$ are flat space-time Gamma matrices related with
curved-space ones by $\Gamma^{\mu}=e^{\mu}_{\ a}\gamma^{a}$. They
form a Clifford algebra in d dimensions, i e, they satisfy the
anti-conmutation relations $\{\gamma^{a},\gamma^{b}\}=-2 \eta^{ab}$,
with $\eta^{00}=-1$.

Under a conformal transformation of the metric of the form :
\begin{eqnarray}
 & g_{\mu\nu}=\Omega^{2}\tilde{g}_{\mu\nu} ,
\end{eqnarray}
the spinor $\psi$ and the Dirac operator transforms as \cite{Gibbons,gibbons-rogatko,splitfermion}
\begin{eqnarray}
\psi &=&\Omega^{-\frac{1}{2}(D-1)}\tilde{\psi} ,\\
\nablasl \psi &=&\Omega^{-\frac{1}{2}(D+1)} \ \tilde{\nablasl}\tilde{\psi}.
\end{eqnarray}
If the line element of a space-time takes the form of a sum of independent components $ds^{2}=ds_{1}^{2}+ds_{2}^{2}$, where $ds_{1}^{2} =g_{uv}(z) dz^u dz ^v$ and $ds_{2}^{2}= g_{mn}(y) dy^m dy ^n$
then Dirac operator $\nablasl$ satisfies a  direct sum decomposition
\beq
\nablasl = \nablasl_z + \nablasl_y.
\eeq
The above mentioned properties of spinors and the Dirac operator in direct sum metrics and conformal related ones, allows a simple treatment of the massless Dirac equations in curved backgrounds. For two
conformally related metrics, the validity of massless Dirac equation in one
implies the the validity of the same equation in the other. Then, the idea is to solve the Dirac equation in the curved space described by an initial metric tensor performing successive conformal transformations
that isolate the metric components that depend of a given variable, and applied successively the direct sum decomposition of Dirac operator, until to obtain an equivalent problem in a space-time of the form $M^{2}\times\Sigma^{D-2}$, where $M^{2}$ is two-dimensional Minkowski space-time and $\Sigma^{D-2}$ is a $(D-2)$-dimensional maximally symmetric manifold, in which the spectrum of massless Dirac operator is known.

The method work as follows \cite{owenjeferson3,mitesis}. First, consider for simplicity a particular class of metrics of the type (\ref{warpedmetric2})
\begin{equation}\label{metricspinorgral}
ds^{2}=\mathfrak{g}_{\mu \nu}dz^{\mu}dz^{\nu}=-A(y^{1})(dy^{0})^{2}+B(y^{1})(dy^{1})^{2}+C(y^{1})d\Sigma_{D-2}^{2}(x^{i}).
\end{equation}
As a basis of one-forms in this space-time we can choose \cite{lopez-ortega}
\begin{eqnarray}\label{vierbeinbasis}
  e^{a} &=& e^{a}(y^{c})=f^{a}_{\ b}(y^{c})dy^{b}, \ \ \ \ a,b=1,2 \\
  e^{i} &=& e^{i}(x^{k})=h^{i}_{\ j}(x^{k})dx^{j}, \ \ \ \ i,j,k=1,...,D-2 .
\end{eqnarray}
This basis is convenient because most of the connection one-forms are equal to zero:  $\omega_{jab}=\omega_{jak}=\omega_{aja}=\omega_{ajk}=0$. This fact allow us to write for the covariant derivatives $\nabla_{a}=\nabla^{(2)}_{a}\otimes\textbf{I}_{(D-2)}, a=0,1$; $\nabla_{j}=\textbf{I}_{(D-2)}\otimes\nabla^{(2)}_{a}, j=2,...,D-1$, where $\textbf{I}_{(D-2)}$ is the identity matrix in $(D-2)$-dimensions \cite{lopez-ortega}.
Performing a conformal rescaling of the original metric (\ref{metricspinorgral}) as $ds^{2}=C(y^{1})d\tilde{s}^{2}$,
where $d\tilde{s}^{2} = -\frac{A}{C}(dy^{0})^{2} + \frac{B}{C}(dy^{1})^{2} + d\Sigma^{2}_{D-2} ,$
then we obtain the equation $\tilde{\nablasl}\tilde{\psi}=\tilde{\Gamma}^{\mu}\tilde{\nabla}_{\mu}\tilde{\psi}=0$, with  $\tilde{\psi}=C^{\frac{(D-1)}{2}}\psi $.

In general, we can find a representation of the Gamma matrices $\Gamma_{\mu}^{(D)}$ in $D$ dimensions as a function of their two-dimensional partners: $\tilde{\Gamma}_{0}^{(2)}, \tilde{\Gamma}_{1}^{(2)}$, and the chirality matrix $\tilde{\bar{\Gamma}}^{(2)}\equiv\tilde{\Gamma}_{5}^{(2)}=\tilde{\Gamma}_{0}^{(2)}\tilde{\Gamma}_{1}^{(2)}$.
If $D$ is even, the Gamma matrices are of dimensionality $2^{\frac{D}{2}} \times 2^{\frac{D}{2}}$, and for odd $D$ they have dimension $2^{\frac{D-1}{2}} \times 2^{\frac{D-1}{2}}$. We use for odd $D$ the following representation for the Gamma matrices
\begin{eqnarray}\label{gamma1}
  \nonumber \tilde{\Gamma}_{0}^{(D)} &=& \tilde{\Gamma}_{0}^{(2)} \otimes \mathbb{I}_{2}\otimes \mathbb{I}_{2}\otimes\mathbb{I}_{2}\otimes \cdot\cdot\cdot=\tilde{\Gamma}_{0}^{(2)} \otimes \mathbb{I}_{2^{\frac{(D-2)}{2}}} \\
  \nonumber \tilde{\Gamma}_{1}^{(D)} &=& \tilde{\Gamma}_{1}^{(2)} \otimes \mathbb{I}_{2}\otimes \mathbb{I}_{2}\otimes\mathbb{I}_{2}\otimes \cdot\cdot\cdot=\tilde{\Gamma}_{1}^{(2)} \otimes \mathbb{I}_{2^{\frac{(D-2)}{2}}} \\
  \nonumber \tilde{\Gamma}_{2}^{(D)} &=& \tilde{\bar{\Gamma}}^{(2)}\otimes \tilde{\Gamma}_{0}^{(2)}\otimes \mathbb{I}_{2}\otimes\mathbb{I}_{2}\otimes \cdot\cdot\cdot \\
  \nonumber \tilde{\Gamma}_{3}^{(D)} &=& \tilde{\bar{\Gamma}}^{(2)}\otimes \tilde{\Gamma}_{1}^{(2)}\otimes \mathbb{I}_{2}\otimes\mathbb{I}_{2}\otimes \cdot\cdot\cdot \\
  \tilde{\Gamma}_{4}^{(D)} &=& \tilde{\bar{\Gamma}}^{(2)}\otimes \tilde{\bar{\Gamma}}^{(2)}\otimes \tilde{\Gamma}_{0}^{(2)}\otimes\mathbb{I}_{2}\otimes \cdot\cdot\cdot \\
  \nonumber \cdot \\
  \nonumber \cdot \\
   \nonumber \cdot \\
   \nonumber \tilde{\Gamma}_{D-1}^{(D)} &=& \tilde{\bar{\Gamma}}^{(2)}\otimes \tilde{\bar{\Gamma}}^{(2)}\otimes\tilde{\bar{\Gamma}}^{(2)}\otimes\cdot\cdot\cdot\otimes\tilde{\Gamma}_{1}^{(2)},
\end{eqnarray}
and for even $D$ we change the last equation above by
 $\tilde{\Gamma}_{D-1}^{(D)}=\tilde{\bar{\Gamma}}^{(2)}\otimes \tilde{\bar{\Gamma}}^{(2)}\otimes\tilde{\bar{\Gamma}}^{(2)}\otimes\cdot\cdot\cdot\otimes\tilde{\Gamma}_{0}^{(2)}$.
It is convenient to define the Gamma matrices $\hat{\tilde{\Gamma}}^{d}$, $d=2,...,D-1$ in the manifold $d\Sigma_{D-2}^{2}$ as:
\begin{eqnarray}\label{gamma2}
  \nonumber \hat{\tilde{\Gamma}}_{2} &=& \tilde{\Gamma}_{0}^{(2)} \otimes \mathbb{I}_{2}\otimes \mathbb{I}_{2}\otimes\mathbb{I}_{2}\otimes \cdot\cdot\cdot \\
  \nonumber \hat{\tilde{\Gamma}}_{3} &=& \tilde{\Gamma}_{1}^{(2)} \otimes \mathbb{I}_{2}\otimes \mathbb{I}_{2}\otimes \cdot\cdot\cdot\\
  \nonumber \hat{\tilde{\Gamma}}_{4} &=& \tilde{\bar{\Gamma}}^{(2)}\otimes \tilde{\Gamma}_{0}^{(2)}\otimes \mathbb{I}_{2}\otimes \cdot\cdot\cdot\\
            \hat{\tilde{\Gamma}}_{5} &=& \tilde{\bar{\Gamma}}^{(2)}\otimes \tilde{\Gamma}_{1}^{(2)}\otimes \mathbb{I}_{2}\otimes \cdot\cdot\cdot \\
  \nonumber \cdot \\
  \nonumber \cdot \\
   \nonumber \cdot \\
   \nonumber \hat{\tilde{\Gamma}}_{D-1} &=& \tilde{\bar{\Gamma}}^{(2)}\otimes \tilde{\bar{\Gamma}}^{(2)}\otimes\tilde{\bar{\Gamma}}^{(2)}\otimes \cdot\cdot\cdot
\end{eqnarray}
Then, we can write the last $D-3$ curved space Gamma matrices in (\ref{gamma1}) as:
\begin{eqnarray}\label{gamma3}
  \nonumber \tilde{\Gamma}_{2}^{(D)} &=& \tilde{\bar{\Gamma}}^{(2)}\otimes \hat{\tilde{\Gamma}}_{2} \\
  \nonumber \tilde{\Gamma}_{3}^{(D)} &=& \bar{\tilde{\Gamma}}^{(2)}\otimes \hat{\tilde{\Gamma}}_{3} \\
  \tilde{\Gamma}_{4}^{(D)} &=& \tilde{\bar{\Gamma}}^{(2)}\otimes \hat{\tilde{\Gamma}}_{4}  \\
  \nonumber \cdot \\
  \nonumber \cdot \\
   \nonumber \cdot \\
   \nonumber \tilde{\Gamma}_{D-1}^{(D)} &=& \tilde{\bar{\Gamma}}^{(2)}\otimes \hat{\tilde{\Gamma}}_{D-1} .
\end{eqnarray}
Also, we have the relations $\tilde{\Gamma}^{0 (D)}=-\tilde{\Gamma}_{0}^{(D)}$, $\tilde{\Gamma}^{\alpha (D)}=\tilde{\Gamma}_{\alpha}^{(D)}, \ \alpha=1,...,D-1$ y $\hat{\tilde{\Gamma}}^{\sigma (D)}=\hat{\tilde{\Gamma}}_{\sigma}^{(D)},\ \sigma=2,...,D-1$.
It is interesting to note that under a conformal transformation of the metric the Gamma matrices remains equal: $\tilde{\Gamma}^{\mu}=\Gamma^{\mu}$.

Using the above representation for the Gamma matrices, the Dirac equation can be written as
\begin{eqnarray}
\left[ \left(
\tilde{\Gamma}^{0}_{(2)}\tilde{\nabla}_{0} +
\tilde{\Gamma}^{1}_{(2)}\tilde{\nabla}_{1} \right) \otimes \mathbb{I}_{(D-2)}
 + \tilde{\bar{\Gamma}}_{(2)} \otimes
\left(\hat{\tilde{\Gamma}}^{i}
\tilde{\nabla}_{i}\right)_{\Sigma_{(D-2)}} \right] \tilde{\psi} = 0
, &
\end{eqnarray}
where $\left(\hat{\tilde{\Gamma}}^{k}\tilde{\nabla}_{k}\right)_{\Sigma_{(D-2)}}$
is the Dirac operator in the maximally symmetric $(D-2)$-dimensional space-time, in which we suppose as known its discrete spectrum consisting in the set of orthogonal eigen-spinors $\tilde{\xi}_{\sigma}^{(\pm)}$ defined by:
\begin{equation}
\left(\hat{\tilde{\Gamma}}^{k}\tilde{\nabla}_{k} \right)_{\Sigma_{D-2}}\tilde{\xi}_{\sigma}^{(\pm)} = \pm \kappa_{\sigma} \tilde{\xi}_{\sigma}^{(\pm)} ,
\end{equation}
being $\sigma$ a set of numbers that label the discrte spectra with eigenvalues $\kappa_{\sigma}=i\lambda_{\sigma}$.
Now expanding the spinor $\tilde{\psi}$ in the basis $\tilde{\xi}_{\sigma}^{(\pm)}$:
\begin{equation}
\tilde{\psi} = \sum_{\sigma} \left(\tilde{ \varphi}_{\sigma}^{(+)} \tilde{\xi}_{\sigma}^{(+)} +\tilde{ \varphi}_{\sigma}^{(-)}\tilde{ \xi}_{\sigma}^{(-)} \right) .
\end{equation}
we can writte the massless Dirac equation in the form
\begin{equation}\label{eqn:2Ddirac}
\left[ \tilde{\Gamma}_{(2)}^{0}\tilde{ \nabla}_{0} + \tilde{\Gamma}_{(2)}^{1}\tilde{ \nabla}_{1}\pm \kappa \tilde{\bar{\Gamma}}_{(2)} \right] \tilde{\varphi}_{\sigma}^{(\pm)} = 0 ,
\end{equation}
In the following we work only with the $+$ sign, because the opposite sign case can be worked in the same form. We also fix the variables in $\mathcal{N}^{2}$ as temporal and radial ones, $(t,r)$. Denoting as $d\tilde{s}_{1}^{\ 2}=-\frac{A(r)}{C(r)}dt^{2} + \frac{B(r)}{C(r)}dr^{2}$ the two-dimensional part of $d\tilde{s}^{2}$ and performing a conformal re-scaling in the form $d\tilde{s}_{1}^{\ 2}=\frac{A}{C} \left[ -dt^{2} +
dr_{*}^{2}\right]$, where $dr_{*}=\sqrt{\frac{B}{A}}dr$ defines the tortoise coordinate we obtain:
\begin{equation}\label{eqn:2Ddirac}
\left [ \gamma^{t}\partial_{t} +  \gamma^{r_{*}}\partial_{r_{*}} + \kappa_{\sigma} \gamma^{5} \sqrt{\frac{A}{C}}\right] \tilde{\varphi}_{\sigma}^{(+)} = 0 ,
\end{equation}
being $\gamma^{t}$ y $\gamma^{r_{*}}$ Dirac matrices and $\gamma^{5}$ the chirality matrix in the two-dimensional Minkowsky space-time with line element $ds^{2}=-dt^{2}+dr_{*}^{2}$.
Choosing for the Dirac matrices the representation:
\begin{equation}
\gamma^{t} = -i\sigma^{3}\ \ \ ,\ \ \
\gamma^{r_{*}}= \sigma^{2} \ \ \ ,\ \ \
\gamma^{5} = (-i\sigma^{3})(\sigma^{2}) = - \sigma^{1},
\end{equation}
where the $\sigma^{i}$ are the Pauli matrices:
\begin{equation}
\sigma^{1}=\left(
\begin{array}{cc}
0 & 1 \\ 1 & 0
\end{array}
\right)\ \ \ ,\ \ \ \sigma^{2}=\left(
\begin{array}{cc}
0 & -i \\ i & 0
\end{array}
\right)\ \ \ ,\ \ \ \sigma^{3}=\left(
\begin{array}{cc}
1 & 0 \\ 0 & -1
\end{array}
\right),
\end{equation}
and writing
\begin{equation}
\tilde{\varphi}_{\sigma}^{(+)} = \left(
\begin{array}{c}
i\zeta_{\sigma}(t,r) \\ \chi_{\sigma}(t,r)
\end{array}
\right) ,
\end{equation}
we obtain the following equations for the components od the Dirac spinor $\tilde{\varphi_{\sigma}}$:
\begin{equation}
i\frac{\partial\zeta_{\sigma}}{\partial t}+\frac{\partial\chi_{\sigma}}{\partial r_{*}}+\Lambda_{\sigma}\chi_{\sigma}=0,
\end{equation}
\begin{equation}
i\frac{\partial\chi_{\sigma}}{\partial t}-\frac{\partial\zeta_{\sigma}}{\partial r_{*}}+\Lambda_{\sigma}\zeta_{\sigma}=0,
\end{equation}
where
\beq \label{lambdaspinor}
\Lambda_{\sigma}(r)= \lambda_{\sigma}\sqrt{\frac{A}{C}} .
\eeq
The above equations can be decoupled to obtain:
\begin{equation}
\frac{\partial^{2}\zeta_{\sigma}}{\partial t^{2}}-\frac{\partial^{2}\zeta_{\sigma}}{\partial r_{*}^{2}}+ V_{+}(r)\zeta_{\sigma}=0,
\label{tevolspinor1}
\end{equation}
\begin{equation}
\frac{\partial^{2}\chi_{\sigma}}{\partial t^{2}}-\frac{\partial^{2}\chi_{\sigma}}{\partial r_{*}^{2}}+ V_{-}(r)\chi_{\sigma}=0,
\end{equation}
where:
\begin{equation}
V_{\pm}=\pm\frac{d\Lambda_{\sigma}}{dr_{*}}+\Lambda_{\sigma}^{2} .
\label{potspinor}
\end{equation}
This last equations describe the time evolution of Dirac perturbations in curved space-times. As the effective potentials $V_{+}$ and $V_{-}$ are super-symmetric in the sense considered by Chandrasekhar in reference \cite{chandra}, then $\zeta_{\sigma}(t,r)$ and $\chi_{\sigma}(t,r)$ will have similar time evolutions and spectra, both quasinormal and scattering. At this point it is interesting to mention that for the spinor solution $\tilde{\varphi}_{\sigma}^{(-)}$, we obtain the same effective potentials. In the following we will work with equation (\ref{tevolspinor1}) and drop the label $(+)$ for the effective potential, defining $V^{(\frac{1}{2})}_{\sigma}(r)\equiv V_{+}(r)$.

\section{Object picture, quasinormal modes and late time tails for fermion perturbations}

\begin{figure}[t]
\begin{center}
\scalebox{0.95}{\includegraphics{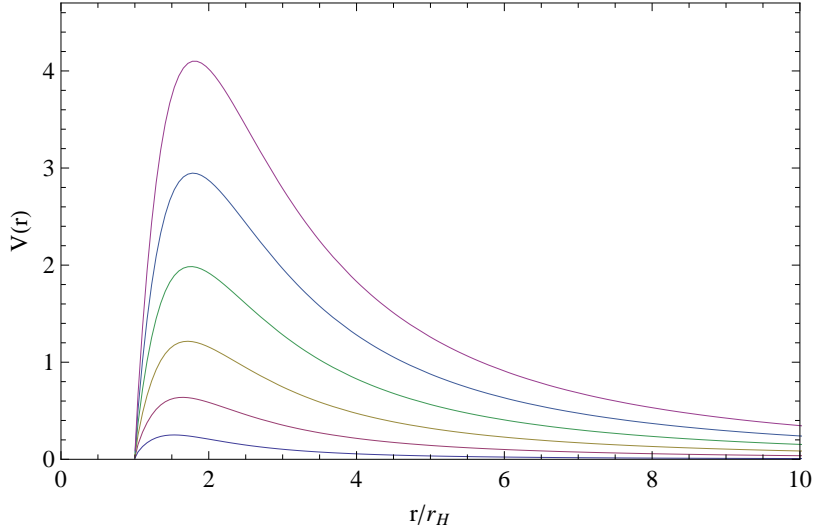}}
\end{center}
\caption{Potential $V(r)$ for Dirac perturbations with $\ell=0$
(bottom) to $\ell=5$(top). The charges parameters of the solution
are $Q_{1}=Q_{2}=Q_{3}=1$ and the dimension is $D=5$ }\label{potencialesdirac}
\end{figure}

The general formalism of the previous section can easily be applied to background space-times of higher dimensional stringy black holes. The line element (\ref{metric}) that describe such systems can be easily written as (\ref{metricspinorgral}) with the identifications
\begin{equation}\label{metricd}
A(r)=h(r)^{-\frac{D-3}{D-2}}f(r), \ \
B(r)=h(r)^{\frac{1}{D-2}}f(r)^{-1}, \ \
C(r)=r^{2}h(r)^{1/D-2},
\end{equation}
and taking \(\mathcal{K}^{n}\) as the $(D-2)$-sphere $\mathcal{S}^{(D-2)}$. In this specific case in which $d\Sigma_{(D-2)}$ is the line element  $d\Omega_{2}^{D-2}$ of the $(D-2)$-sphere we have for the eigenvalues of the Dirac equations $\kappa_{\ell}=\pm i\lambda_{\ell}$, where $\lambda_{\ell}=\ell+\frac{D-2}{2}, \ \ell=0,1,2,...$, being $\ell$ the angular multipole number \cite{Camporesi}.

Taking the above elements into account, we obtain for the effective potencial $V(r)$ the general expression:
\begin{equation}
 V(r)=\frac{\lambda_\ell f^{\frac{3}{2}}}{2 r h}\left[ \frac{f'}{f}-\frac{h'}{h}-\frac{2}{r}\left(1-\frac{\lambda_{\ell}}{\sqrt{f}}\right)\right].
\label{pot1}
\end{equation}
Figure (\ref{potencialesdirac}) shows the effective potential for different multipole numbers $\ell$ in the case of five dimensional stringy black holes with  $Q_1=Q_2=Q_3=1$. The form of the effective potential is similar for other dimensions and setups for compactification charges. As we can see, the effective potential have the form of a definite-positive potential barrier, tending to zero at infinity. This character of the effective potential assures the stability of the stringy black holes under such spinor perturbations.

\begin{figure}[t]
\begin{center}
\scalebox{0.70}{\includegraphics{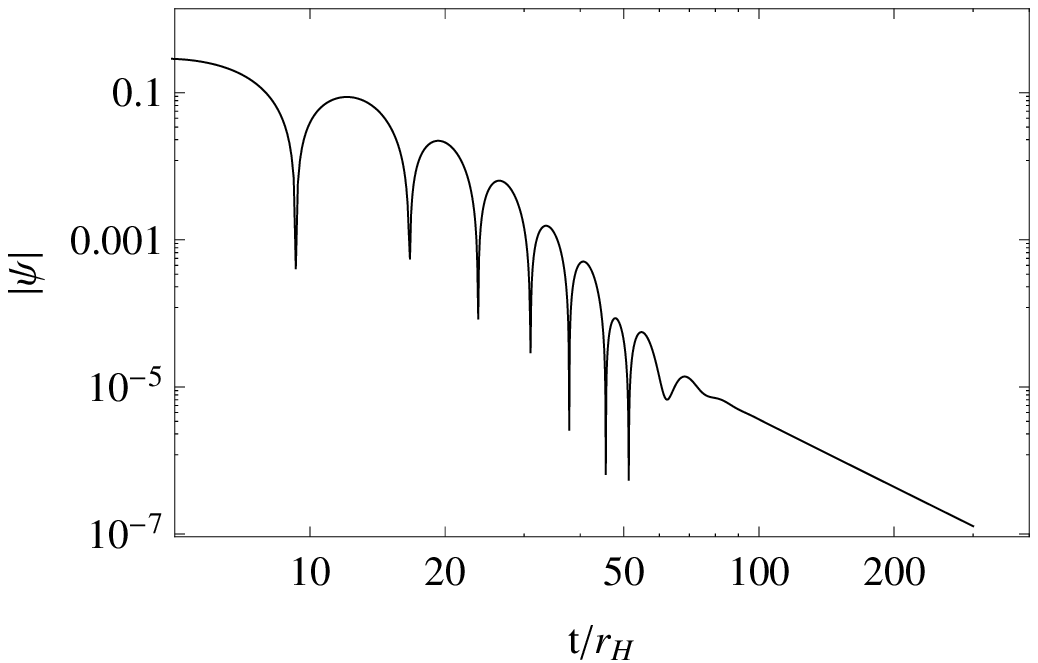}}
\vspace{0.35cm}
\scalebox{0.70}{\includegraphics{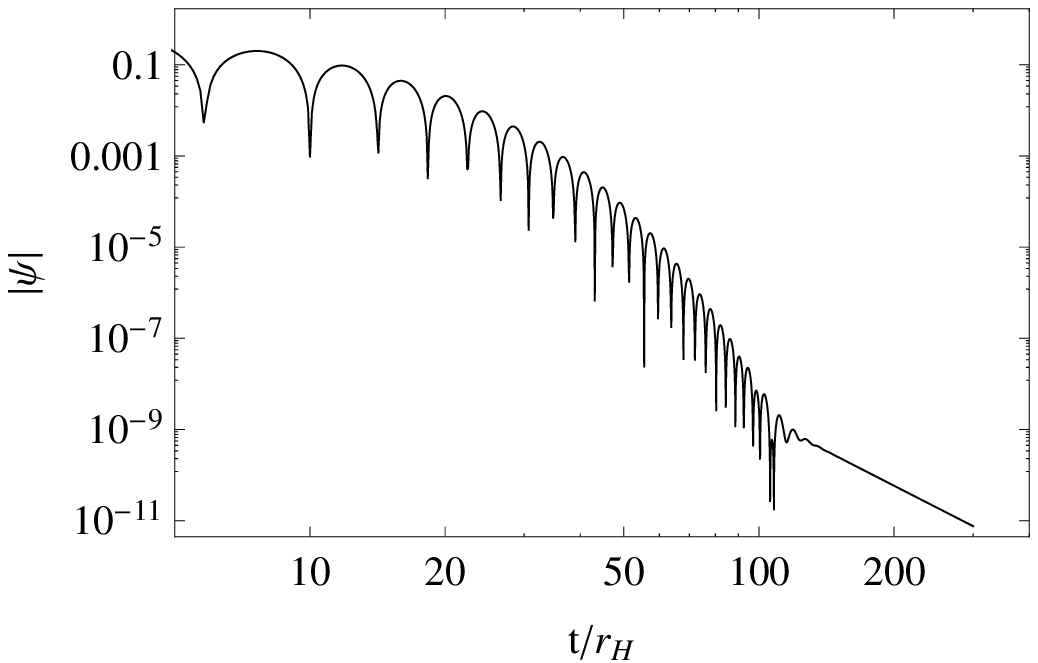}}
\scalebox{0.70}{\includegraphics{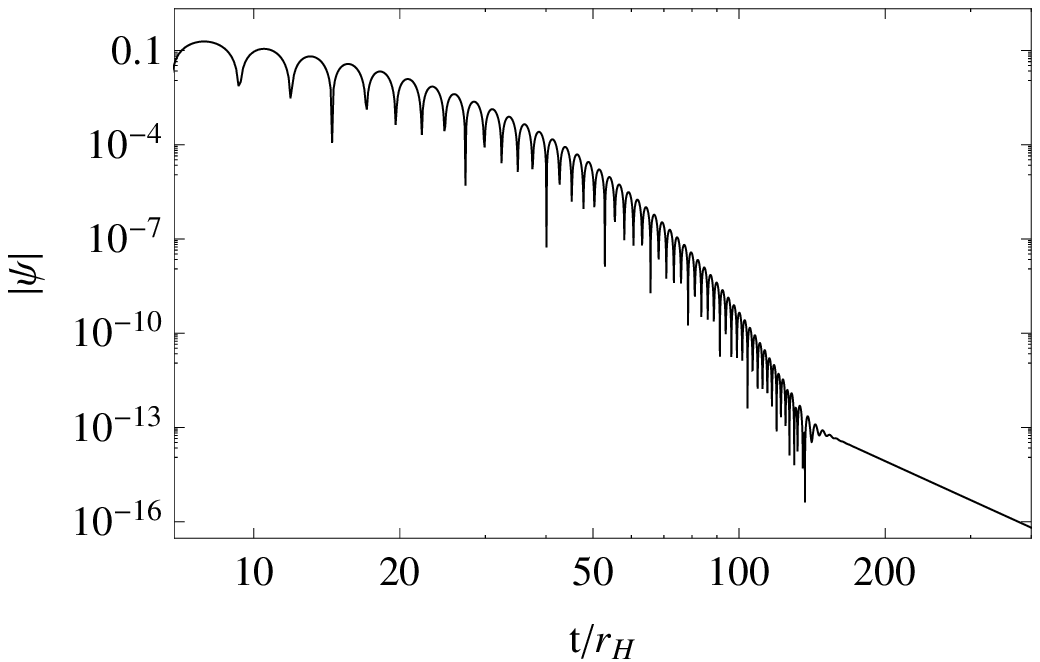}}
\vspace{0.35cm}
\scalebox{0.70}{\includegraphics{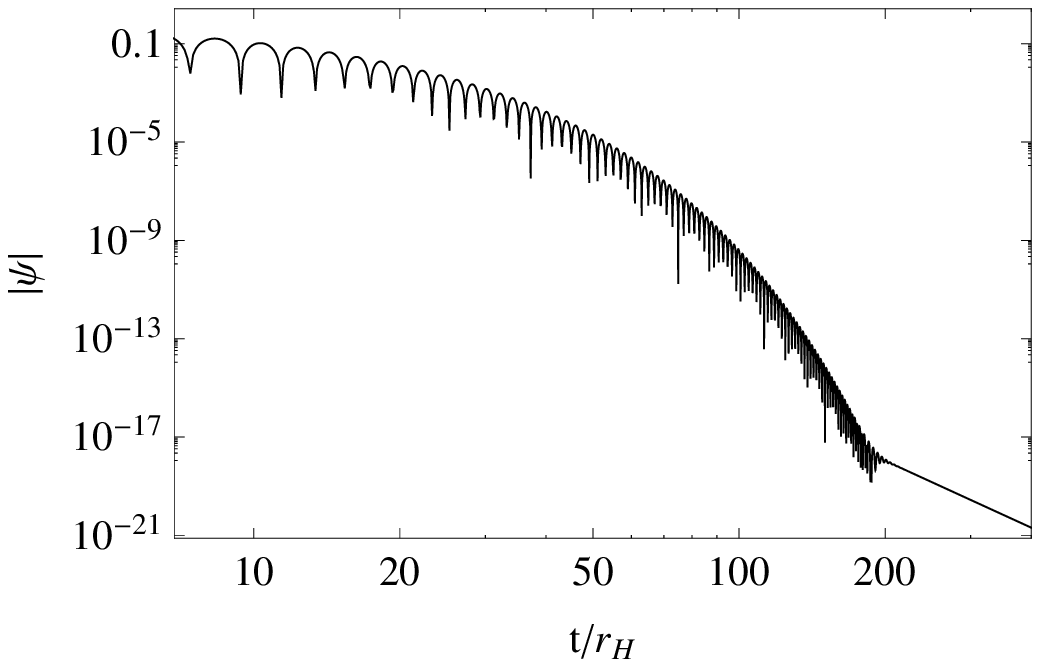}}
\end{center}
\caption{\it Logarithmic plots of the
time-domain evolution of $\ell=0$ (top left), $\ell=1$ (top right), $\ell=2$ (bottom left) and $\ell=3$ (bottom right)
massless Dirac perturbations in $(4+1)$-stringy black holes with charges $Q_{1}=0, Q_{2}=Q_{3}=1$ (top) and
$Q_{1}=Q_{2}=Q_{3}=1$ (bottom).} \label{perfil1}
\end{figure}

\begin{figure}[t]
\begin{center}
\scalebox{0.70}{\includegraphics{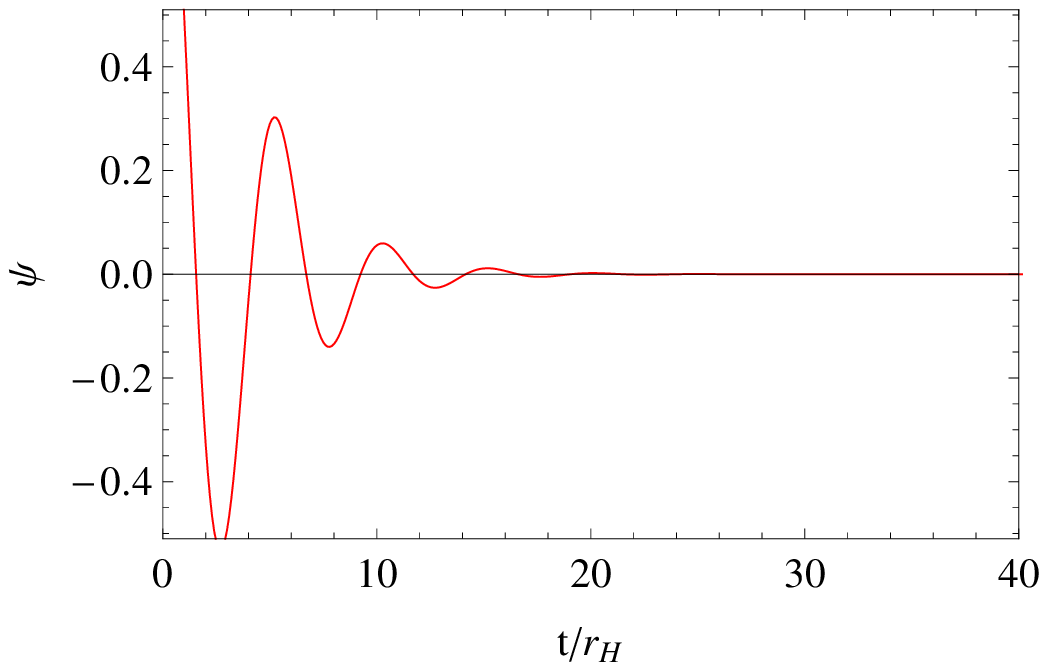}}
\vspace{0.35cm}
\scalebox{0.70}{\includegraphics{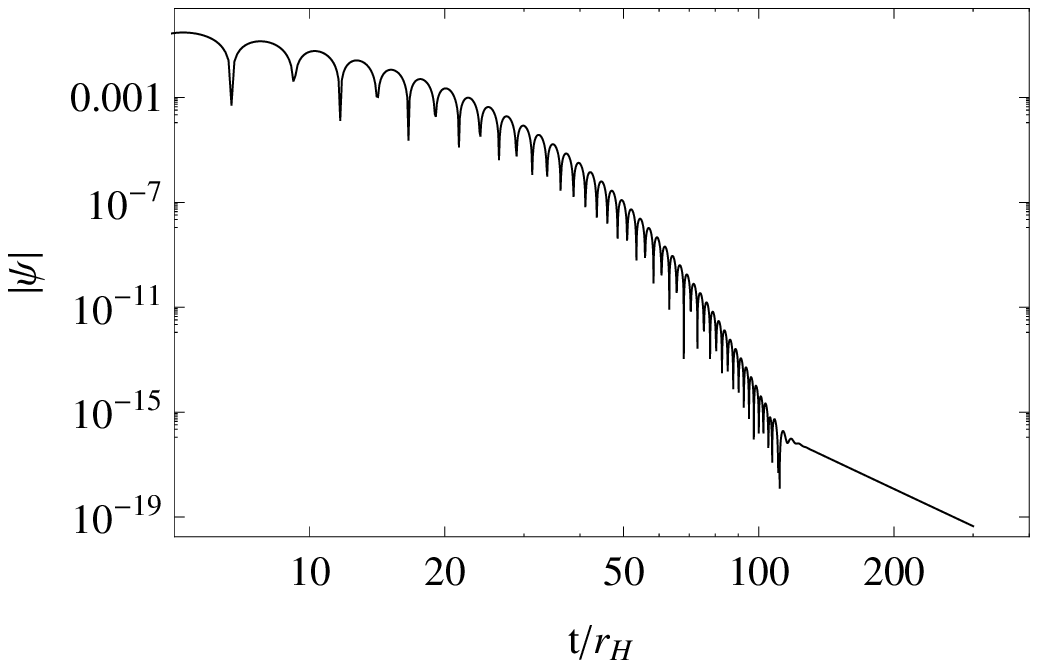}}
\end{center}
\caption{\it Normal (left) and logarithmic (right) plots of the
time-domain evolution of $\ell=1$ massless Dirac perturbations in $(5+1)$-stringy black holes with charges
$Q_{1}=Q_{2}=1$.} \label{perfil2}
\end{figure}

\begin{figure}[t]
\begin{center}
\scalebox{0.70}{\includegraphics{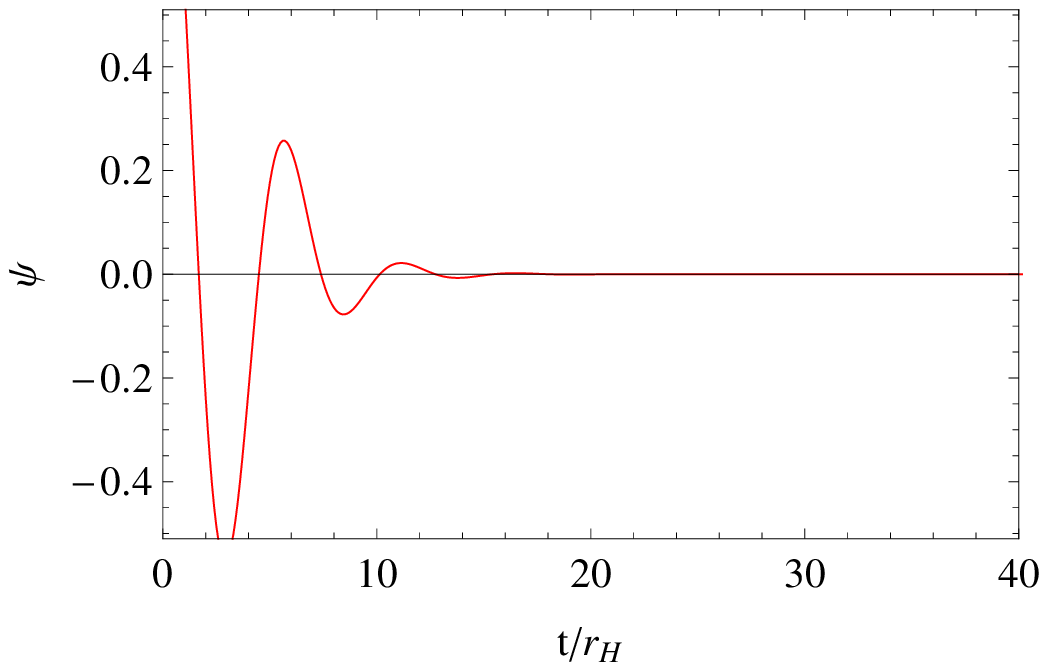}}
\vspace{0.35cm}
\scalebox{0.70}{\includegraphics{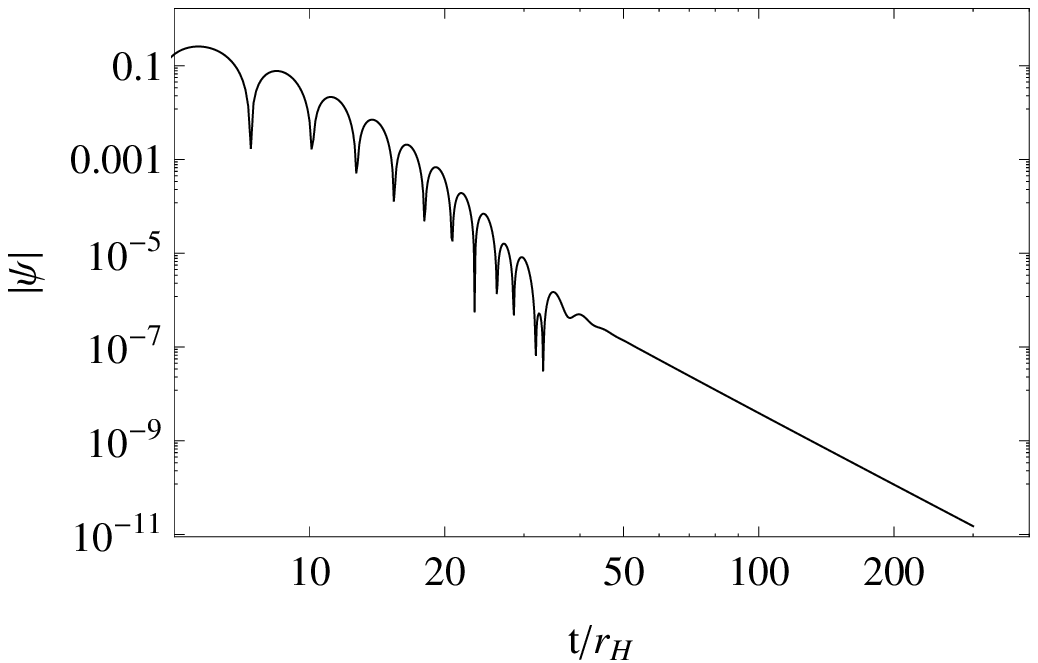}}
\scalebox{0.70}{\includegraphics{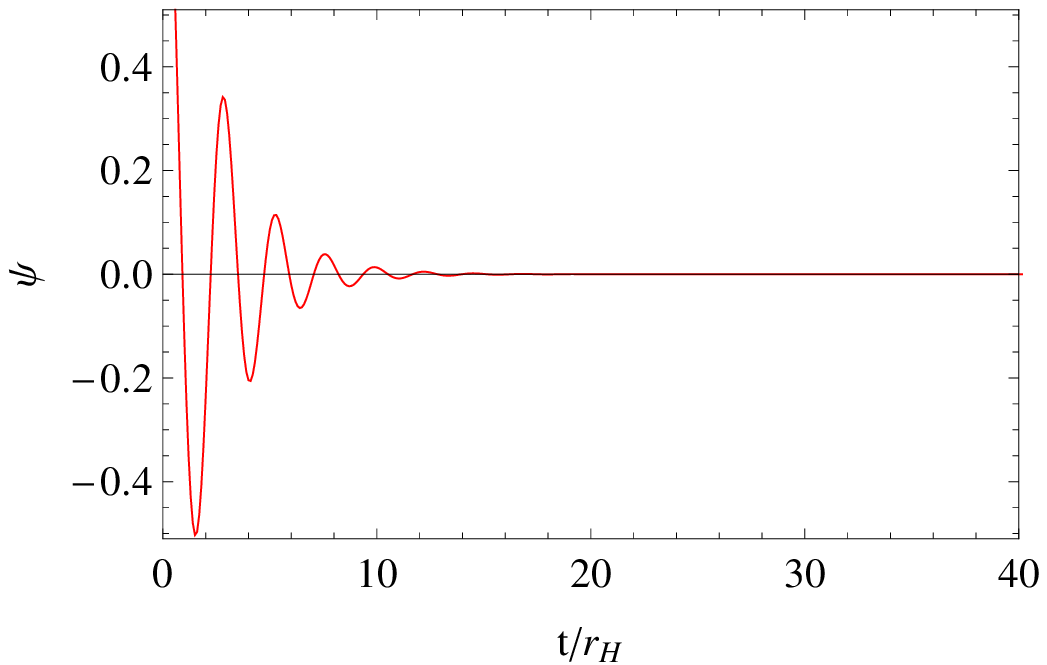}}
\vspace{0.35cm}
\scalebox{0.70}{\includegraphics{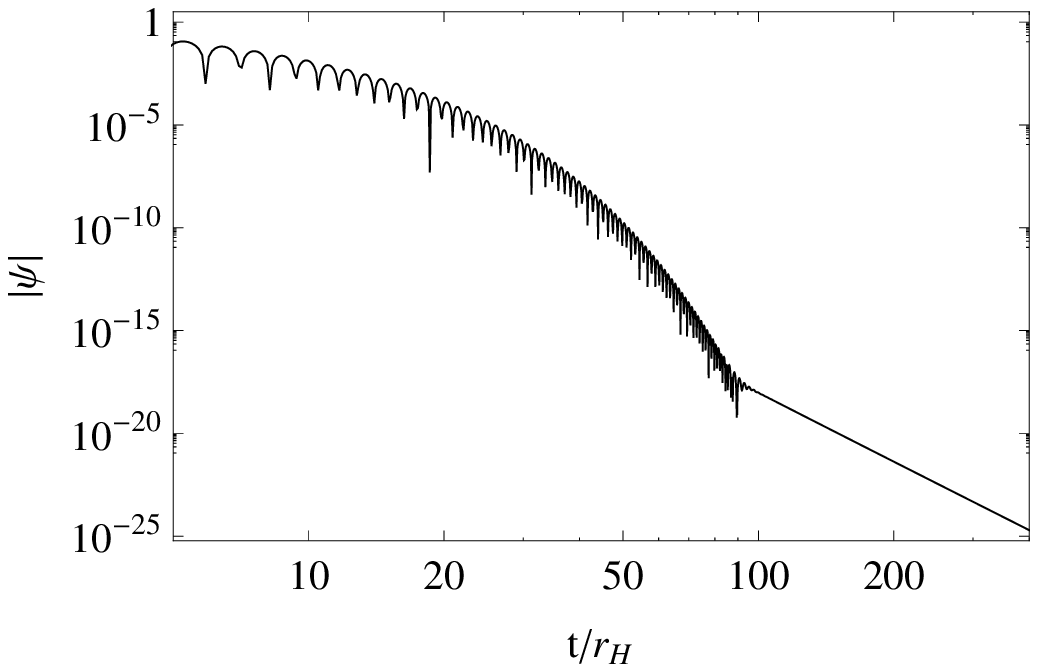}}
\end{center}
\caption{\it Normal (left) and logarithmic (right) plots of the
time-domain evolution of $\ell=0$ (top) and $\ell=3$ (bottom)
massless Dirac perturbations in $(6+1)$-stringy black holes with charges
$Q_{1}=Q_{2}=1$.} \label{perfil3}
\end{figure}

\begin{figure}[t]
\begin{center}
\scalebox{0.70}{\includegraphics{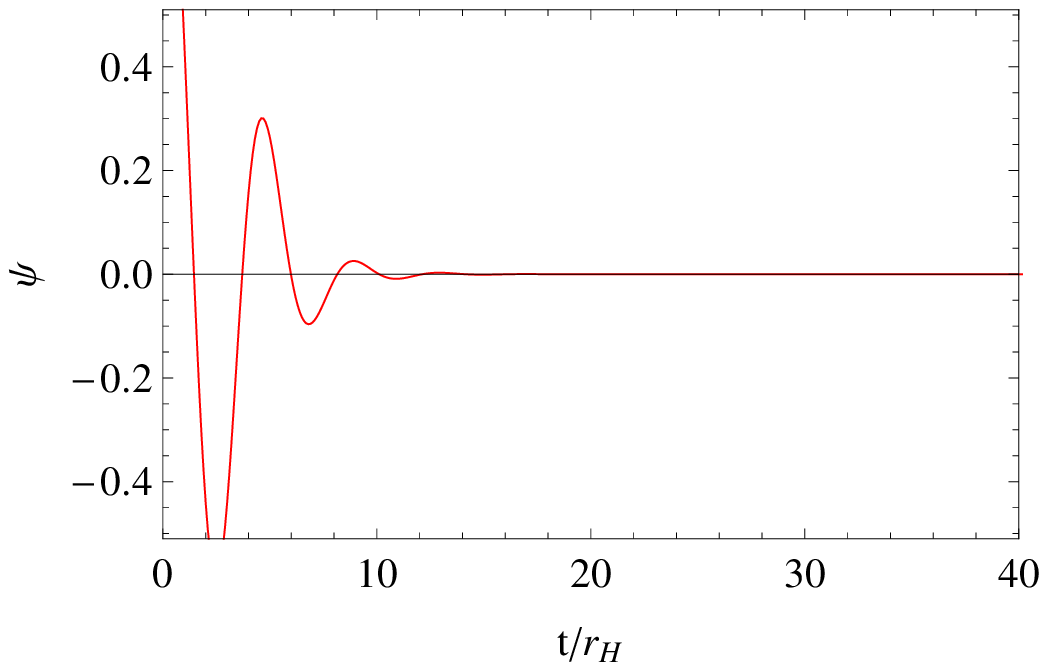}}
\vspace{0.35cm}
\scalebox{0.70}{\includegraphics{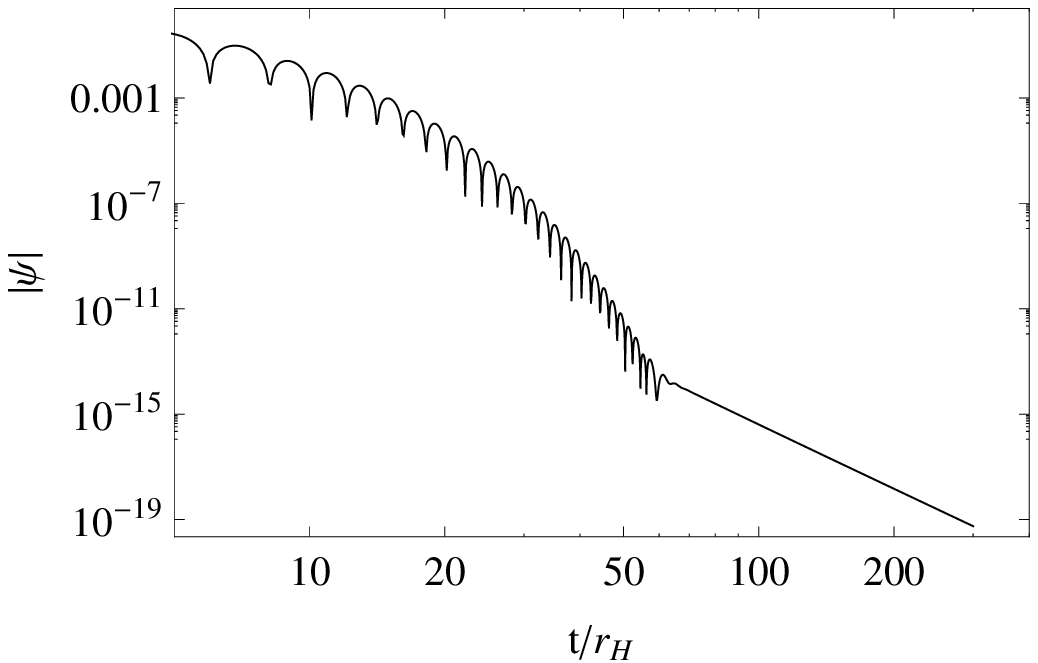}}
\scalebox{0.70}{\includegraphics{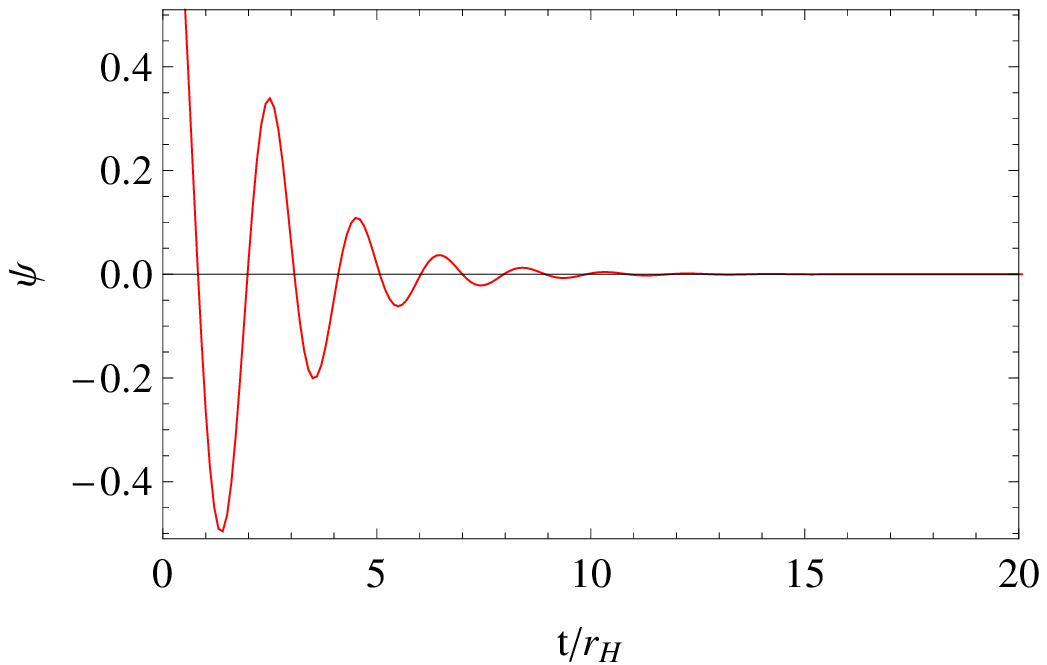}}
\vspace{0.35cm}
\scalebox{0.70}{\includegraphics{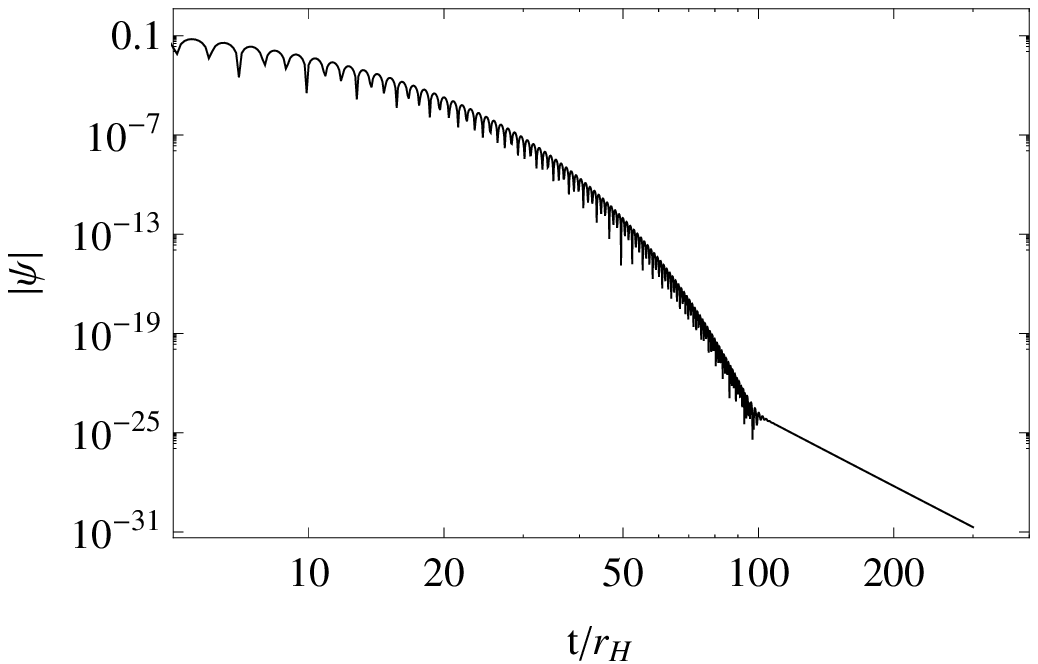}}
\end{center}
\caption{\it Normal (left) and logarithmic (right) plots of the
time-domain evolution of $\ell=0$ (top) and $\ell=3$ (bottom)
massless Dirac perturbations in $(7+1)$-stringy black holes with charges
$Q_{1}=Q_{2}=1$.} \label{perfil4}
\end{figure}

\begin{figure}[t]
\begin{center}
\scalebox{0.70}{\includegraphics{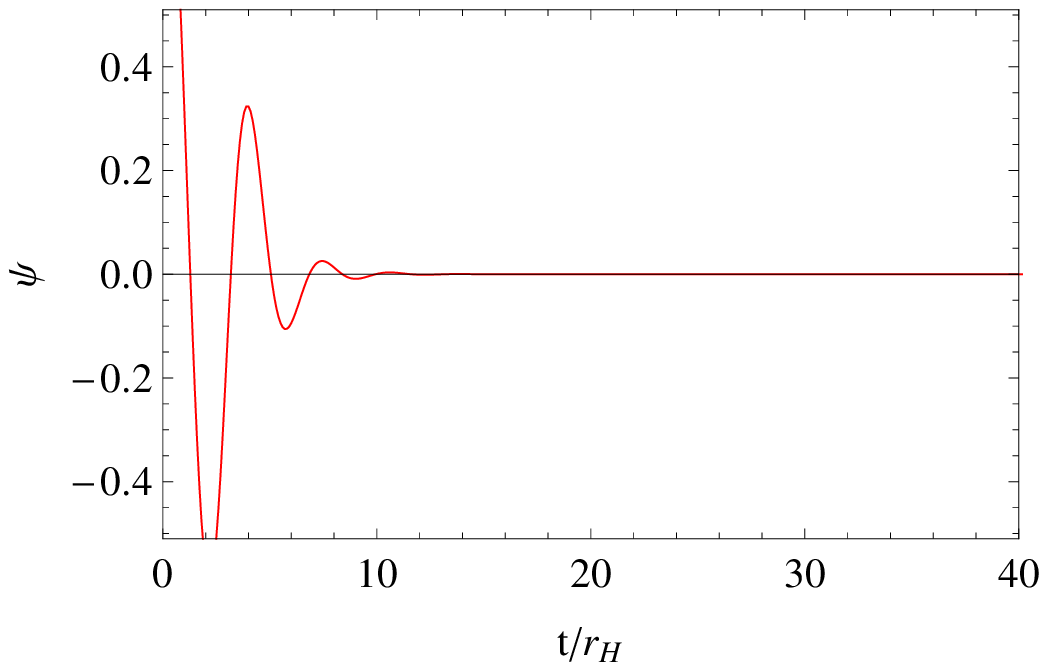}}
\vspace{0.35cm}
\scalebox{0.70}{\includegraphics{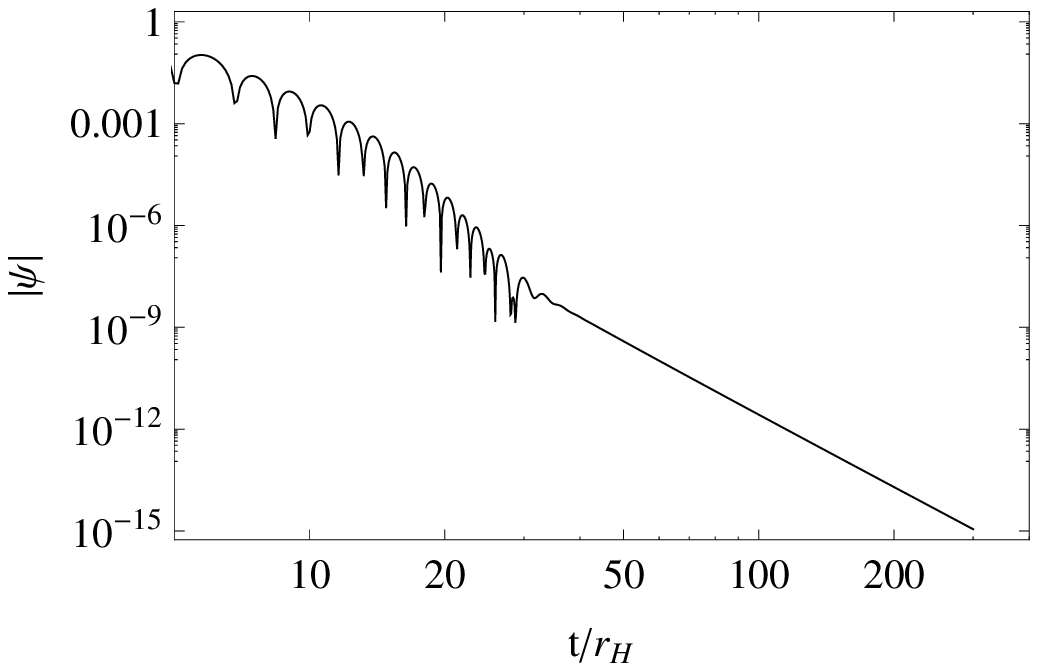}}
\scalebox{0.70}{\includegraphics{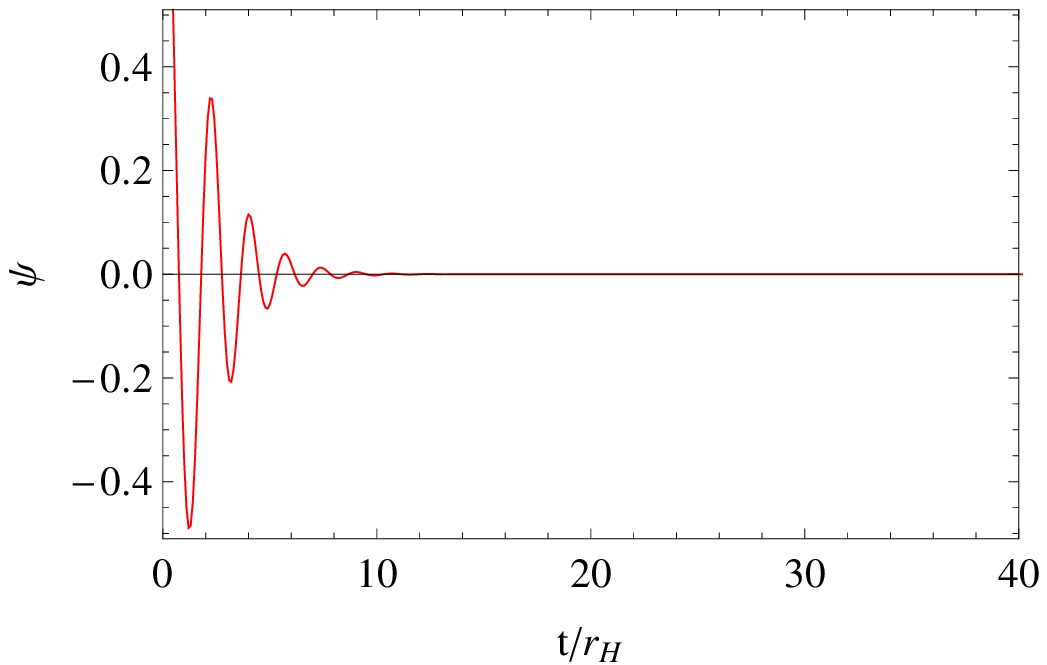}}
\vspace{0.35cm}
\scalebox{0.70}{\includegraphics{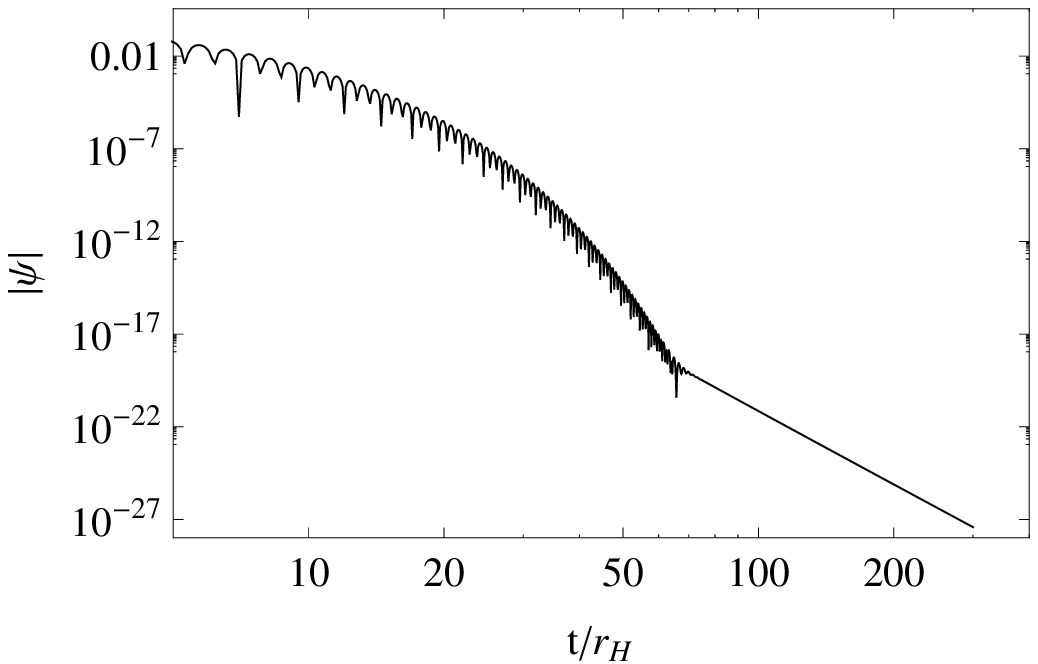}}
\end{center}
\caption{\it Normal (left) and logarithmic (right) plots of the
time-domain evolution of $\ell=0$ (top) and $\ell=3$ (bottom)
massless Dirac perturbations in $(8+1)$-stringy black holes with charges
$Q_{1}=Q_{2}=1$.} \label{perfil5}
\end{figure}
To integrate the equation (\ref{tevolspinor1}) numerically we use the characteristic integration technique developed by Gundlach, Price and Pullin \cite{gpp}. The obtained results can be observed as the time-domain profiles  showed in Figures (\ref{perfil1}) to (\ref{perfil5}). In such profiles $r=3 r_{H}$  and  the time is measured in units of black hole event horizon.

The first thing that can be observed from this profiles is that the time evolution of Dirac perturbations outside $D$-dimensional stringy black holes follows the usual dynamics for fields in other black hole and brane backgrounds \cite{Abdalla:2008te}. After a first transient stage strongly dependent on the initial conditions and the point where the wave profile is observed, we see the characteristic exponential damping of the perturbations associated with the quasinormal ringing, followed by power law tails at asymptotically late times. This picture is independent of the chosen values for the compactification charges, and the dimension of the background space-time. It its also appreciated that the duration of the quasinormal phase is shorter as the space-time dimension is larger. This behavior again is of the same qualitative character for different setups of compactification charges.

To compute the quasinormal frequencies that dominated at intermediate times, we assume for the function $\zeta_{\ell}(t,r)$ in equation (\ref{tevolspinor1}) the time dependence:
\begin{eqnarray}
\ \ \ \ \ \ \ \  \zeta_{\ell}(t,r)=Z_{\ell}(r)\exp(-i\om_{\ell}t)
\end{eqnarray}
Then, the function $Z_{\ell}(r)$  satisfy the Schrodinger-type equation:
\begin{equation}\label{finaleq}
\frac{d^{2} Z_{\ell}}{d r_{*}^{2}}+\left[\om^{2}- V(r)\right] Z_{\ell}(r)=0
\end{equation}
The quasinormal modes are solutions of the equation (\ref{finaleq})
with the specific boundary conditions requiring pure out-going waves
at spatial infinity and pure in-coming waves on the event horizon.
Thus no waves come from infinity or the event horizon.

For the higher dimensional stringy black holes considered in this work, the quasinormal frequencies were computed using two different
methods. The first method uses directly the numerical data obtained previously, and fit this data by superposition of damping exponents. This numerical fitting scheme, known as Prony method, allow us to obtain very accurate results for the fundamental and sometimes the first overtones \cite{berti3,zhidenkothesis}. For higher overtones is very difficult to be implemented, because we need to do a fitting with a great number of exponentials, and also we need to know very well the particular time in which quasinormal ringing begins, a difficult point to be solved in general.  For this reason we only determined the quasinormal frequencies, using this method, for the fundamental overtone.

The second method that we employed is a semi-analytical approach to solve equation
(\ref{finaleq}) with the required boundary conditions, based in a
WKB-type approximation, that can give accurate values of the lowest
( that is longer lived ) quasinormal frequencies, and  was used in
several papers for the determination of quasinormal frequencies in a
variety of systems \cite{splitfermion,Abdalla:2008te,shutz-will,iyer-will,konoplya1,konoplya2,WKB6papers,WKB6papers1,WKB6papers2,WKB6papers3,WKB6papers4,WKB6papers5}.

Tables I to X in the Appendix at the end of this report presents the values obtained for the quasinormal frequencies with some multipole numbers $\ell$ for stringy black holes at different dimensions and different set of charges. As it is observed, the sixth order WKB approach gives results in agreement with those obtained by fitting the numerical integration data using Prony technique.

As it is expected the oscillation frequency increases for higher multipole and fixed overtone numbers. Increasing $\ell$ the magnitude of the negative imaginary part of the fundamental overtone ($n=0$) increases while for higher overtones the opposite situation arises. For a fixed angular number $\ell$, the real part of the oscillation frequencies decreases as the overtone number increases, and the magnitude of the imaginary part increases. Then, modes with higher overtone numbers decay faster than low-lying ones.

As we expected for stability, all quasinormal frequencies calculated in this work have a well defined finite negative imaginary part.
 \begin{figure}[htb!]
\begin{center}
\scalebox{0.40}{\includegraphics{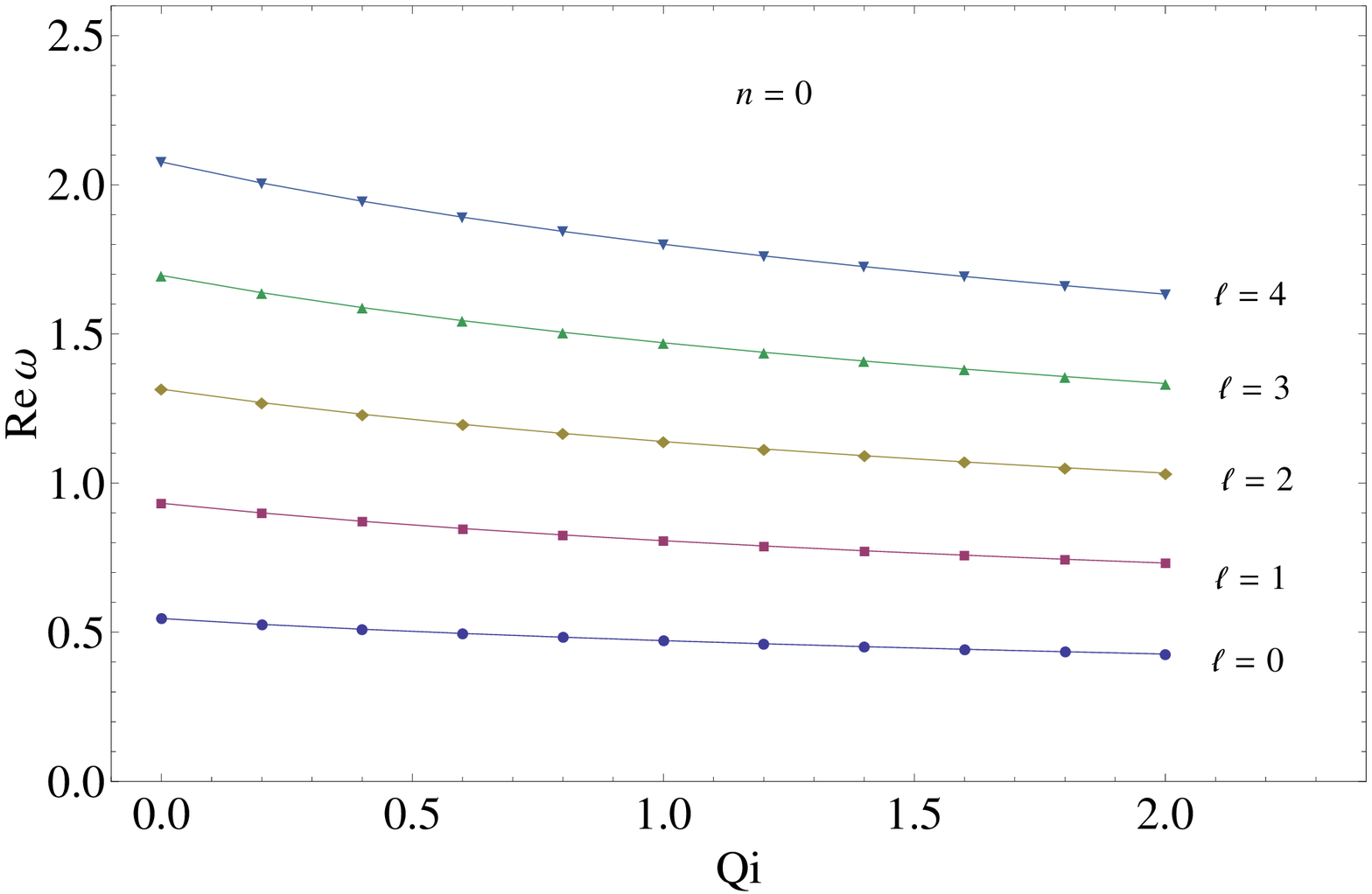}}
\scalebox{0.40}{\includegraphics{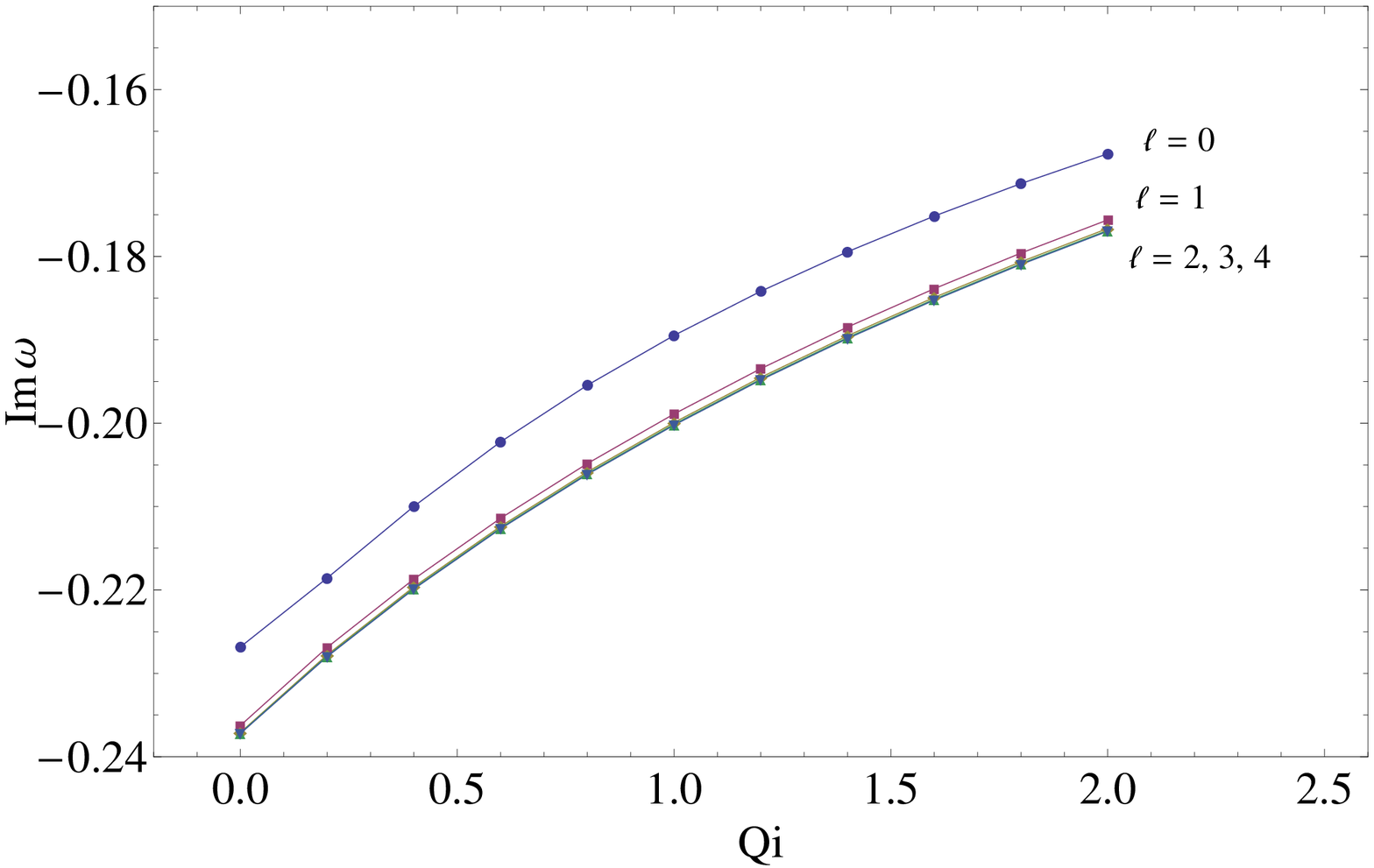}}
\end{center}
\caption{\it Dependence upon charge $Q_i$, $i=1,2,3$ of the real (left) and imaginary (right) parts of the massless Dirac quasinormal frecuencies of five dimensional stringly black holes with the other two charges fixed $Q_{j}=1$, $j\neq i$. The results correspond to the first overtone for multipolar number from $\ell=0$ to $\ell=4$.} \label{stringy5dwvsq}
\end{figure}

Figure (\ref{stringy5dwvsq}) shows the dependence of the quasinormal modes with compactification charges in the case of five-dimensional stringy black holes. Again the results for higher dimensions are exhibit the same qualitative behaviour. As one of the charges increases, the frequency and decay factor of the perturbation decreases. For fixed values of the charges, the oscillation frequency increases as multipole numbers increases, for fixed overtone number. As an interesting fact we see that the decay factor of quasinormal modes with $n=0$ reaches quickly a saturation value for higher multipole numbers. The situation is different for higher overtone numbers, as the decay factor of the quasinormal oscillations for fixed $n$ decreases as $\ell$ increases. Again, the qualitative behaviour described is similar for other dimensions.

\begin{figure}[t]
\begin{center}
\scalebox{0.45}{\includegraphics{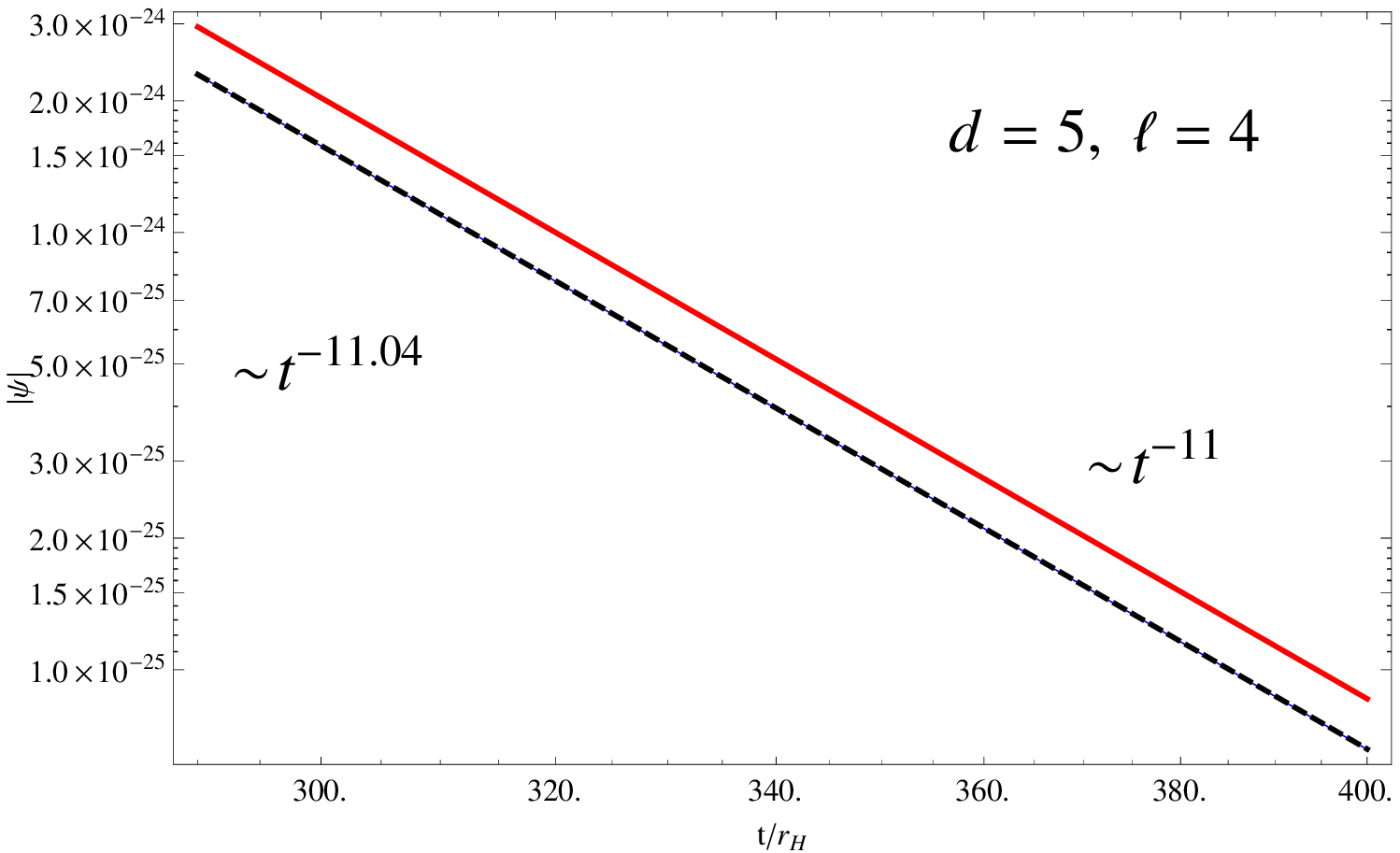}}
\vspace{0.35cm}
\scalebox{0.39}{\includegraphics{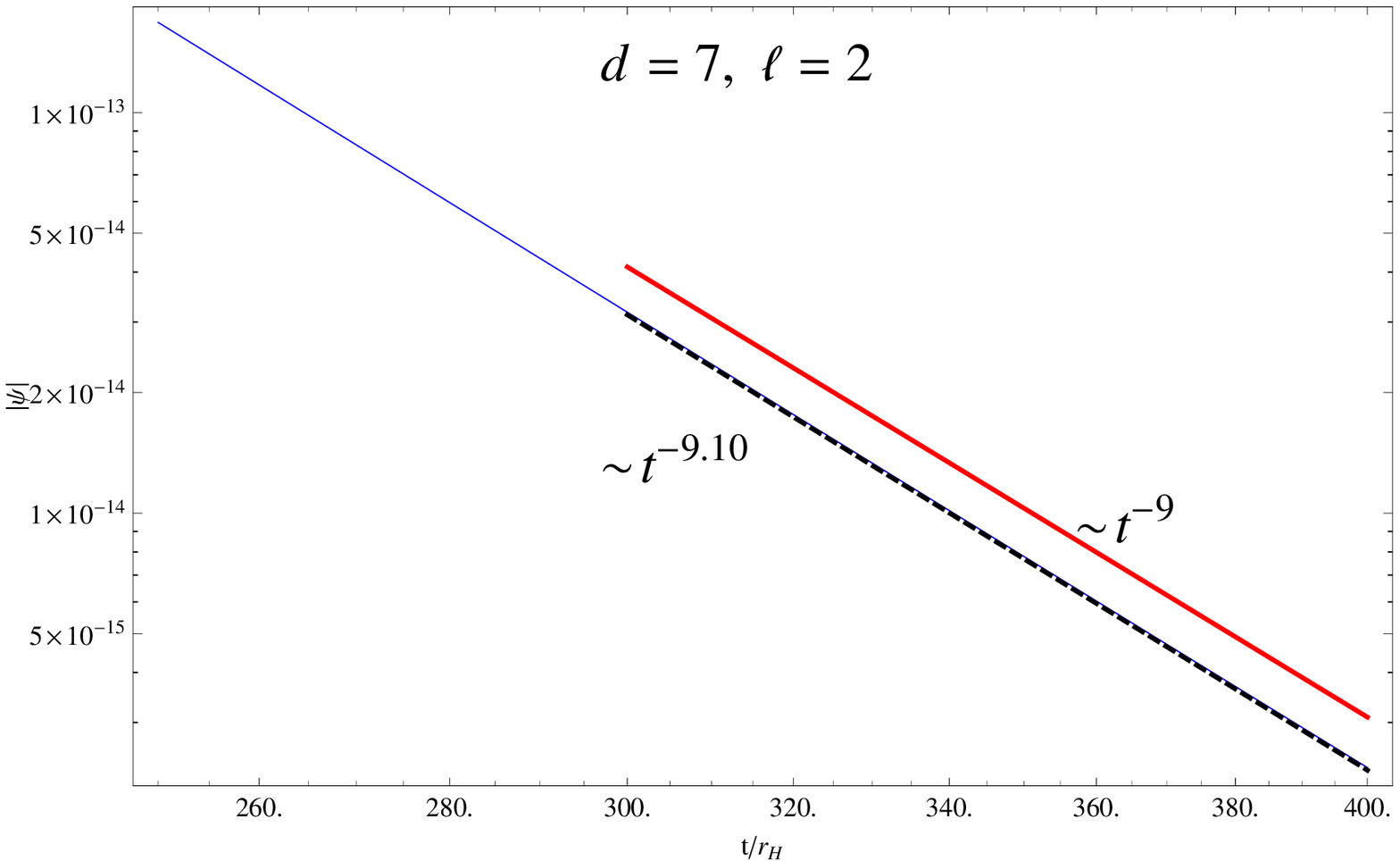}}
\scalebox{0.39}{\includegraphics{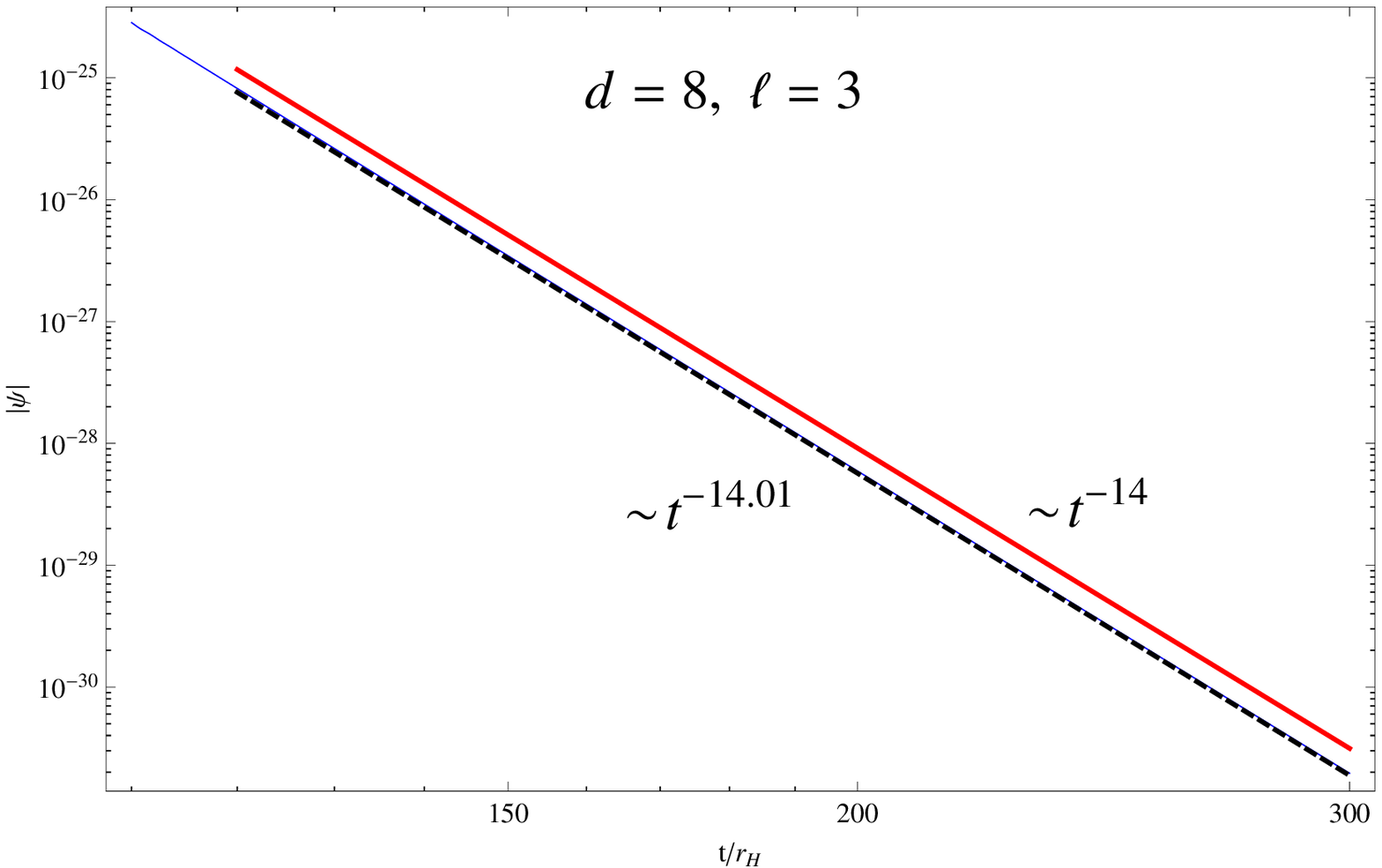}}
\vspace{0.35cm}
\scalebox{0.39}{\includegraphics{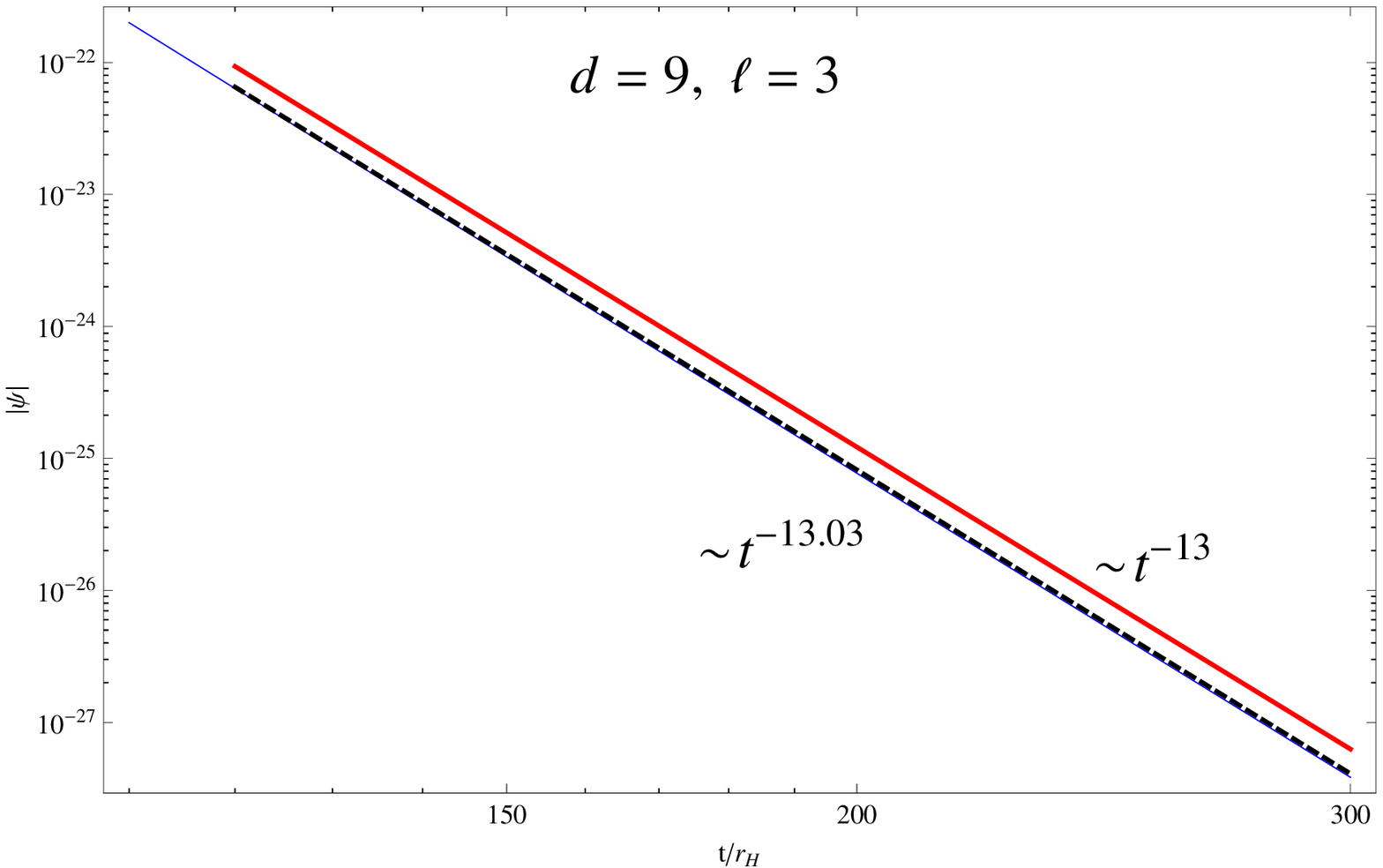}}
\end{center}
\caption{{\it{Late time tails in profiles of time evolution of fermion perturbations in stringy black holes. The power-law coefficients
were estimated from numerical data represented in the dotted line.
The full red line is the possible analytical result.The parameters in the figure are $D=5$, $\ell=4$, $Q_{1}=0,Q_{2}=Q_{3}=1$ (top left);  $D=7$, $\ell=2$, $Q_{1}=Q_{2}=Q_{3}=1$ (top right); $D=8$, $\ell=3$, $Q_{1}=Q_{2}=Q_{3}=1$ (bottom left) and $D=9$, $\ell=3$, $Q_{1}=Q_{2}=Q_{3}=1$. }}} \label{tail1}
\end{figure}

Another important point to rise here is the relaxation of the fermion disturbance outside the black hole at very late times \cite{Price,burko}. To study the late-time behavior, we numerically fit the profile data obtained in the appropriate region of the time domain, to extract the power law exponents that describe the relaxation.

Figure (\ref{tail1}) show some representative examples of the results obtained, for different higher dimensional stringy black holes and given values of compactification charges. In general, all our numerical results indicate that the decay of massless Dirac fields shows a time dependance in the form of a power law, whose exponent changes not only with the multipole number $\ell$, but also with the dimension $D$ of the background space-time.
Our numerical study shows that the power law relaxation factors depends stronlgy of the odd or even nature of the space-time dimension $D$. Stringy black holes of odd dimensions relax at asymptotically late times following a power law proportional to $t^{-(2\ell+D-2)}$, whereas for the case of even dimension the falloff is more quickly, following a time evolution proportional to $t^{-(2\ell+3D-8)}$ for $D=6$, and to $t^{-(2\ell+3D-16)}$ for $D=8$.

It is important to stress that a similar dependance was analytically obtained for the case of scalar perturbations in space-times of Schwarzschild-Tangherlini \cite{cardoso1}. In this case it was proved that the time evolution at asymptotically late times in odd dimensional space-times is independent of the presence of the black hole, because of the Green function of the Laplace-Beltrami operator in asymptotically flat backgrounds had always the same tail. For even dimensions, the result is strongly dependent on the presence of the black hole and, as in the four dimensional case, the tails is determined by the asymptotic behaviour of the effective potential at larger values of the tortoise coordinate.

In reference \cite{cardoso1} tails proportional to $t^{-(2\ell+3D-8)}$ were obtained, for any even dimension $D$. However, as our results shows, fermion perturbations decay following this law only for $D=6$, whereas the disturbance falloff outside $8$-dimensional stringy black holes is more slowly. Thus, it seems that the fermion or boson character of the test field have influence in the late falloff of the perturbations in even higher dimensions.

However, we remark at this point that this dependances are only a result consistent with our numerical data. A simple analytical argument to support this late time behaviours do not exist, in contrast to the case for boson fields, in which the general form of the effective potential is suitable to expand for large values of the tortoise coordinate \cite{cardoso1,ching1,ching2} and then extract the above power law behaviors directly from this asymptotic expansion. The problem related with the analytical determination of the decay factors for fermion perturbations in higher dimensional stringy black holes remains open.

\section{Large angular momentum limit.}

An interesting case arises for large angular multipole numbers because in this limit the first order WKB approximation becomes exact and we can obtain an analytical result for the quasinormal frequencies. As an interesting fact we should mention that in this limit, for all the higher dimensional stringy back holes considered in this report, the effective potentials are given by the same general expression
\begin{equation}
V(r)=\ell^{2}\Delta(r)
\label{pot2}
\end{equation}
where $\Delta(r)=\frac{f(r)}{r^2 h(r)}$. Then the first order WKB approximation gives for the quasinormal frequencies in this limit the result:
\begin{equation}\label{wkb_large_l}
\omega^{2}=\ell^{2}\Delta(r_{m})-i\ell(n+\frac{1}{2})\sqrt{-2\Xi(r)|_{r=r_{m}}},
\end{equation}
with $\Xi(r)\equiv\frac{d^{2}\Delta(r)}{dr_{*}^{2}}$ is given by
\begin{eqnarray*}
  \Xi(r)&=& \frac{1}{4r^{4}f^{\frac{7}{2}}}\left\{9r^{2}f^{2}h'^{2}+rhf\left[h'(18f-11rf')-4rfh''\right] \right.
  \nonumber \\
  &&\left.\,+\ 2h^{2}\left[12f^{2}+r^{2}f'^{2}+2rf(r f''-5f')\right]\right\}
\end{eqnarray*}
In (\ref{wkb_large_l}) $r_{m}$ the point at which the asymptotic effective potential (\ref{pot2}) reach its peak.  This value can be determined as the maximum root of the equation
\begin{equation}
-f(r)\left[2h(r)+r\frac{dh(r)}{dr}\right]+rh(r)\frac{df(r)}{dr}=0.
\end{equation}
In the particular case of equal charges $Q_{i}=Q$, the above general equation results
\begin{equation}
\left[(3D-11)Q+D-1\right]r^{3-D}r_{H}^{D+3}-2(D-4)Qr^{6-2D}r_{H}^{2D}-2r_{H}^{6}=0,\ \ \ \ \ D=5.
\end{equation}
and
\begin{equation}
\left[2(D-4)Q+D-1\right]r^{3-D}r_{H}^{D+3}-(D-5)Qr^{6-2D}r_{H}^{2D}-2r_{H}^{6}=0,\ \ \ \ \ 6\leq D \geq9.
\end{equation}
In five dimensions we easily obtain the result:
\begin{equation}
r_{m}=r_{H}\sqrt{1+Q+\sqrt{1+Q+Q^{2}}}
\label{root5}
\end{equation}
whereas for other space-time dimensions we have:
\begin{equation}
r_{m}=r_{H}\sqrt[3]{\frac{5}{4}+Q+\sqrt{\frac{25}{16}+Q(Q+2)}},  \ \ \ \ \ \ \ \ \ \ D=6
\label{root6}
\end{equation}
\begin{equation}
r_{m}=r_{H}\sqrt[4]{\frac{3(Q+1)}{2}+\sqrt{\frac{9(Q^{2}+1)}{4}+4Q}},  \ \ \ \ \ D=7
\label{root7}
\end{equation}
\begin{equation}
r_{m}=r_{H}\sqrt[5]{\frac{7}{4}+2Q+\sqrt{\frac{49}{16}+\frac{Q}{2}(8Q+13)}},  \ \ \ \ \ D=8
\label{root8}
\end{equation}
and
\begin{equation}
r_{m}=r_{H}\sqrt[6]{\frac{5}{2}Q+2+\sqrt{4+\frac{Q}{2}(25Q+38)}},  \ \ \ \ \ D=9
\label{root9}
\end{equation}

\section{Concluding remarks}

In this paper we studied the evolution of massless Dirac disturbances outside space-times corresponding to higher dimensional black hole solutions coming from intersecting branes in string theories.

As in other black hole and brane backgrounds, the time evolution of fermion perturbations id divided in three epochs. After the initial transient stage, the evolution is dominated by quasinormal oscillations, and at late times by power-low falloffs. We computed the quasinormal frequencies for different values of background dimensions and compactification charges using two different approaches, a semi-analytical 6th-order WKB formula and Prony fitting of the numerical data obtained from the characteristic integration of the evolution equations. Both methods gives very close numerical results for the quasinormal frequencies. The results for the dependence of the quasinormal frequencies with charges appears to be universal for all values of this parameter and all dimensions.

We also computed the decay factors for the late time relaxation of fermion perturbations, showing that for odd space-time dimensions fermion fields relaxes in the same way as scalar waves in uncharged backgrounds. However, the late time falloff of the Dirac fluctuatiosn for even background dimensions is strongly dependent on the dimension itself, a result that contrast with the scalar case.

Some extensions to this work are interesting. First, it would be an important question to determine analytically the late time decay factors, taking into account that for potentials typical of fermion fields do not exist a simple analytical argument to approach this problem, as in the case of boson fields. Also, the study of the evolution of boson perturbations: scalar, electromagnetic and gravitational ones in this higher dimensional space-times would be an interesting subject to be addressed. With the solution of some of those important problems we will be involved in future reports.

\section*{Acknowledgments}
This work has been supported by FAPESP (\emph{Funda\c c\~ao de
Amparo \`a Pesquisa do Estado de S\~ao Paulo}), Brazil and the University of Cienfuegos, Cuba.
One of the authors (OPFP) express his gratitude to Professor Elcio Abdalla from the Department of Mathematical Physics at University of S\~{a}o Paulo for the support and useful conversations during a research stay at his group, where this report was completed.
\newpage
\section*{Appendix: Numerical Results for quasinormal frequencies.}

In this Appendix we will show the tables containing the numerical
results for the calculations of some quasinormal modes of stringy
black holes in different dimensions for some chosen values of the charges.

\begin{table}[htb!]
\begin{center}
  {\begin{tabular}{|c|c|c|c|}
      \hline
      \hline
       $\ell$ & $n$ & $6^{th}$ order WKB  & Prony  \\
      \hline
      $0$ & $0$  & $0.4349 -0.1712i$    &  $ 0.4346 -0.1709i $ \\
      \hline
  $1$ & $0$ &  $0.7445-0.1795i$     & $ 0.7443-0.1792i $ \\
      \hline
  $2$ & $0$ &  $1.0518-0.1806 i$    &  $ 1.0518-0.1806 i $ \\
      \hline
 $2$ & $1$  & $ 1.0073-0.5489i$ & $ - $ \\
     \hline
 $3$ & $0$  & $ 1.3571-0.1809i$ & $ 1.3571-0.1809 $ \\
      \hline
$3$ & $1$   &  $ 1.3224-0.5469i$    & $ - $ \\
     \hline
$3$ & $2$   & $1.2540-0.9263i $ & - \\
      \hline
$4$ &  $0$  & $ 1.6623-0.1810i $    & $ 1.6623-0.1810i  $ \\
      \hline
 $4$ & $1$  & $1.6335-0.5458i$  & $-$ \\
      \hline
$4$ & $2$   &  $1.5769-0.9196i$ & - \\
     \hline
$4$ & $3$   & $1.4942-1.3083i$  & - \\
      \hline
       \hline
   \end{tabular}\label{fdirac8}}
   \caption{\it Massless Dirac
quasinormal frequencies \(\omega r_{H}\) for $\ell=0$ to $\ell=4$.
The charges are chosen as $Q1=0.5$, $Q2=1$, $Q3=1.8$ and the spacetime dimension is $D=5$. }
\end{center}
      \end{table}

\newpage
\begin{table}[htb!]
\begin{center}
  {\begin{tabular}{|c|c|c|c|}
      \hline
      \hline
       $\ell$ & $n$ & $6^{th}$ order WKB  & Prony  \\
      \hline
      $0$ & $0$  & $0.5101-0.2063i$ &  $ 0.5098-0.2059i $ \\
      \hline
  $1$ & $0$ &  $0.8702-0.2152i$     & $0.8700-0.2151i $ \\
      \hline
  $2$ & $0$ &  $1.2270-0.2161i$     &  $1.2270-0.2161i $ \\
      \hline
 $2$ & $1$  & $ 1.1789-0.6579i$ & $- $ \\
     \hline
 $3$ & $0$  & $ 1.5829-0.2163i$ & $1.5829-0.2163i $ \\
      \hline
$3$ & $1$   &  $ 1.5445-0.6547i$    &$- $ \\
     \hline
$3$ & $2$   & $1.4720-1.1106i $ & - \\
      \hline
$4$ &  $0$  & $ 1.9379-0.2163i $    &$  1.9379-0.2163i  $ \\
      \hline
 $4$ & $1$  & $1.9063-0.6530i$  & $-$ \\
      \hline
$4$ & $2$   &  $1.8452-1.1014i$ &$ - $ \\
     \hline
$4$ & $3$   &$ 1.7590-1.5689i$  & $ - $ \\
      \hline
       \hline
   \end{tabular}\label{fdirac9}}
   \caption{\it Massless Dirac
quasinormal frequencies \(\omega r_{H}\) for $\ell=0$ to $\ell=4$.
The charges are chosen as $Q1=0$, $Q2=1$, $Q3=1$ and the spacetime dimension is $D=5$. }
\end{center}
      \end{table}
\newpage
\begin{table}[htb!]
\begin{center}
  {\begin{tabular}{|c|c|c|c|}
      \hline
      \hline
       $\ell$ 	& $n$ &  $6^{th}$ order WKB  & Prony  \\
      \hline
	$0$ 	& $0$  & $0.8545	-0.3048i$    &  $ 0.8547	-0.3048i $ \\
      \hline
	$1$ 	& $0$ &   $1.2830	-0.3344i$     & $ 1.2831	-0.3344i $ \\
      \hline
	$2$ 	& $0$ &    $1.7245	-0.3365i$     &  $ 1.7245	-0.3364i $ \\
      \hline
	$2$ 	& $1$  &   $1.6107	-1.0275i$  & $ - $ \\
     \hline
	$3$ 	& $0$  &   $2.1665	-0.3363i$ & $ 2.1665	-0.3363i $ \\
      \hline
	$3$ 	& $1$   &   $2.0736	-1.0214i$    & $ - $ \\
     \hline
	$3$ 	& $2$   &   $1.8904	-1.7441i$  & $ - $ \\
      \hline
	$4$ 	&  $0$  &   $2.6069	-0.3364i$    & $2.6069	-0.3364i$ \\
      \hline
	$4$ 	& $1$  &  $2.5292	-1.0175i$  & $ - $ \\
      \hline
	$4$ 	& $2$   &    $2.3748	-1.7245i$  & $ - $ \\
     \hline
	$4$ 	& $3$   &   $2.1457	-2.4771i$ & $ - $ \\
     \hline
	$5$ 	& $0$   &   $3.0462	-0.3365i$   & $ 3.0462	-0.3365i$ \\
     \hline
	$5$ 	& $1$   &   $2.9797	-1.0155i$   & $ - $ \\
     \hline
	$5$ 	& $2$   &   $2.8468	-1.7130i$   & $ - $ \\
     \hline
	$5$ 	& $3$   &   $2.6486	-2.4431i$   & $ - $ \\
      \hline
	$5$ 	& $4$   &   $2.3878	-3.2220i$   & $ - $ \\
      \hline
      \hline
   \end{tabular}\label{fdirac10}}
   \caption{\it Massless Dirac
quasinormal frequencies \(\omega r_{H}\) for $\ell=0$ to $\ell=5$.
The charges are chosen as $Q1=Q2=Q3=1$ and the spacetime dimension is $D=5$. }
\end{center}
\end{table}

\newpage
\begin{table}[htb!]
\begin{center}
  {\begin{tabular}{|c|c|c|c|}
      \hline
      \hline
       $\ell$ 	& $n$ &  $6^{th}$ order WKB  & Prony  \\
      \hline
	$0$ 	& $0$  & $0.9753	-0.3387i$    &  $ 0.9755	-0.3387i $ \\
      \hline
	$1$ 	& $0$ &   $1.4500	-0.3844i$     & $ 1.4501	-0.3844i $ \\
      \hline
	$2$ 	& $0$ &    $1.9427	-0.3893i$     &  $ 1.9427	-0.3893i$ \\
      \hline
	$2$ 	& $1$  &   $1.8208	-1.1924i$  & $ - $ \\
     \hline

	$3$ 	& $0$  &   $2.4381	-0.3893i$ & $2.4381	-0.3893i $ \\
      \hline
	$3$ 	& $1$   &   $2.3379	-1.1858i$    & $ - $ \\
     \hline

	$3$ 	& $2$   &   $2.1448	-2.0382i$  & $ - $ \\
      \hline

	$4$ 	&  $0$  &   $2.9323	-0.3892i$    & $ 2.9323	-0.3892i $ \\
      \hline
	$4$ 	& $1$  &  $2.8483	-1.1802i$  & $ - $ \\
      \hline
	$4$ 	& $2$   &    $2.6839	-2.0100i$  & $ - $ \\
     \hline
	$4$ 	& $3$   &   $2.4479	-2.9088i$ & $ - $ \\
\hline
	$5$ 	& $0$   &   $3.4257	-0.3892i$   & $ 3.4257	-0.3892i $ \\
\hline
	$5$ 	& $1$   &   $3.3535	-1.1768i$   & $ - $ \\
\hline
	$5$ 	& $2$   &   $3.2111	-1.9925i$   & $ - $ \\
\hline
	$5$ 	& $3$   &   $3.0037	-2.8579i$   & $ - $ \\
      \hline
	$5$ 	& $4$   &   $2.7404	-3.7965i$   & $ - $ \\
      \hline
      \hline
   \end{tabular}\label{fdirac11}}
   \caption{\it Massless Dirac
quasinormal frequencies \(\omega r_{H}\) for $\ell=0$ to $\ell=5$.
The charges are chosen as $Q1=0$, $Q2=1$ and the spacetime dimension is $D=6$. }
\end{center}
\end{table}

\newpage
\begin{table}[htb!]
\begin{center}
  {\begin{tabular}{|c|c|c|c|}
      \hline
      \hline
       $\ell$ 	& $n$ &  $6^{th}$ order WKB  & Prony \\
      \hline
	$0$ 	& $0$  & $1.2488 -0.3986i$    &  $ 1.2490 -0.3982i $ \\
      \hline
	$1$ 	& $0$ &  $1.7023 -0.4570i$      & $ 1.7024 -0.4570i $ \\
      \hline
	$2$ 	& $0$ &   $2.2068 -0.4575i$      &  $ 2.2068 -0.4575i $ \\
      \hline
	$2$ 	& $1$  & $2.0037 -1.4019 i$   & $ - $ \\
     \hline
	$3$ 	& $0$  & $2.7164 -0.4547 i$  & $ 2.7164 -0.4547 i $ \\
      \hline
	$3$ 	& $1$   &  $2.5460 -1.3868i$     & $ - $ \\
     \hline
	$3$ 	& $2$   & $2.1981 -2.3835i$   & $ - $ \\
      \hline
	$4$ 	&  $0$  &  $3.2225 -0.4539i$     & $ 3.2225 -0.4539i $ \\
      \hline
	$4$ 	& $1$   & $3.0795 -1.3759i$   & $ - $ \\
      \hline
	$4$ 	& $2$   &  $ 2.7861 -2.3427i$   & $ - $ \\
     \hline
	$4$ 	& $3$   &  $ 2.3271 -3.3962i$  & $ - $ \\
\hline
	$5$ 	& $0$   &  $3.7264 -0.4539i$    & $ 3.7264 -0.4539i $ \\
\hline
	$5$ 	& $1$   &  $3.6035 -1.3712i$   & $ - $ \\
\hline
	$5$ 	& $2$   &  $3.3514 -2.3191i$    & $ - $ \\
\hline
	$5$ 	& $3$   &  $2.9592 -3.3271i$   & $ - $ \\
      \hline
	$5$ 	& $4$   &  $2.4175 -4.4380i$    & $ - $ \\
      \hline
      \hline
   \end{tabular}\label{fdirac12}}
   \caption{\it Massless Dirac quasinormal frequencies \(\omega r_{H}\) for $\ell=0$ to $\ell=5$.
The charges are chosen as $Q1=Q2=1$ and the spacetime dimension is $D=7$.}
\end{center}
    \end{table}

\newpage
\begin{table}[htb!]
\begin{center}
  {\begin{tabular}{|c|c|c|c|}
      \hline
      \hline
       $\ell$ 	& $n$ &  $6^{th}$ order WKB  & Prony  \\
      \hline
	$0$ 	& $0$  & $1.4028 -0.4138i$    &  $ 1.4029 -0.4138i $ \\
      \hline
	$1$ 	& $0$ &  $1.8827 -0.5020i$      & $ 1.8829 -0.5019i $ \\
      \hline
	$2$ 	& $0$ &   $2.4272 -0.5092i$      &  $ 2.4272 -0.5092i $ \\
      \hline
	$2$ 	& $1$  & $ 2.2240 -1.5616i$   & $ - $ \\
     \hline

	$3$ 	& $0$  & $2.9827 -0.5071i$  & $2.9827 -0.5071i $ \\
      \hline
	$3$ 	& $1$   &  $ 2.8090 -1.5510i$     & $ - $ \\
     \hline

	$3$ 	& $2$   & $2.4626 -2.6868i$   & $ - $ \\
      \hline

	$4$ 	&  $0$  &  $3.5362 -0.5061i$     & $ 3.5362 -0.5061i$ \\
      \hline
	$4$ 	& $1$   & $3.3890 -1.5387i$   & $ - $ \\
      \hline
	$4$ 	& $2$   &  $3.0927 -2.6365i$   & $ - $ \\
     \hline
	$4$ 	& $3$   &  $2.6443 -3.8594i$  & $ - $ \\
\hline
	$5$ 	& $0$   &  $4.0878 -0.5058i$    & $ 4.0878 -0.5058i$ \\
\hline
	$5$ 	& $1$   &  $3.9608, -1.5318i$   & $ - $ \\
\hline
	$5$ 	& $2$   &  $3.7039 -2.6041i$    & $ - $ \\
\hline
	$5$ 	& $3$   &  $ 3.3140 -3.7652i$   & $ - $ \\
      \hline
	$5$ 	& $4$   &  $2.7947 -5.0715i$    & $ - $ \\
      \hline
      \hline
   \end{tabular}\label{fdirac13}}
   \caption{\it Massless Dirac quasinormal frequencies \(\omega r_{H}\) for $\ell=0$ to $\ell=5$.
The charges are chosen as $Q1=0$, $Q2=1$ and the spacetime dimension is $D=7$.}
\end{center}
      \end{table}

\newpage
\begin{table}[htb!]
\begin{center}
  {\begin{tabular}{|c|c|c|c|}
      \hline
      \hline
       $\ell$ 	& $n$ &  $6^{th}$ order WKB  & Prony  \\
      \hline
	$0$ 	& $0$  & $1.6785	-0.4942i$    &  $ 1.6786	-0.4942i $ \\
      \hline
	$1$ 	& $0$ &  $2.1200	-0.5864i$      & $ 2.1200	-0.5862i $ \\
      \hline
	$2$ 	& $0$ &   $2.6741	-0.5801i$      &  $ 2.6741	-0.5801i$ \\
      \hline
	$2$ 	& $1$  & $2.3620	-1.7865i$   & $ - $ \\
     \hline

	$3$ 	& $0$  & $3.2420	-0.5700i$  & $ 3.2420	-0.5700i $ \\
      \hline
	$3$ 	& $1$   &  $2.9703	-1.7521i$     & $ - $ \\
     \hline

	$3$ 	& $2$   & $2.4081	-3.0249i$   & $ - $ \\
      \hline

	$4$ 	&  $0$  &  $3.8019	-0.5665i$     & $ 3.8019	-0.5665i $ \\
      \hline
	$4$ 	& $1$   & $3.5750	-1.7228i$   & $ - $ \\
      \hline
	$4$ 	& $2$   &  $3.0967	-2.9485i$   & $ - $ \\
     \hline
	$4$ 	& $3$   &  $2.3157	-4.2988i$  & $ - $ \\
\hline
	$5$ 	& $0$   &  $4.3571	-0.5657i$    & $ 4.3571	-0.5657i $ \\
\hline
	$5$ 	& $1$   &  $4.1629	-1.7103i$   & $ - $ \\
\hline
	$5$ 	& $2$   &  $3.7532	-2.8989i$    & $ - $ \\
\hline
	$5$ 	& $3$   &  $3.0894	-4.1785i$   & $ - $ \\
      \hline
	$5$ 	& $4$   &  $2.1364	-5.6218i$    & $ - $ \\
      \hline
      \hline
   \end{tabular}\label{fdirac14}}
   \caption{\it Massless Dirac quasinormal frequencies \(\omega r_{H}\) for $\ell=0$ to $\ell=5$.
The charges are chosen as $Q1=Q2=1$ and the spacetime dimension is $D=8$.}
\end{center}
      \end{table}

\newpage
\begin{table}[htb!]
\begin{center}
  {\begin{tabular}{|c|c|c|c|}
      \hline
      \hline
       $\ell$ 	& $n$ &  $6^{th}$ order WKB  & Prony  \\
      \hline
	$0$ 	& $0$  & $1.9005	-0.5183i$    &  $ 1.9007	-0.5182i $ \\
      \hline
	$1$ 	& $0$ &  $2.3570	-0.6525i$      & $ 2.3572	-0.6525i $ \\
      \hline
	$2$ 	& $0$ &   $2.9553	-0.6538i$      &  $ 2.9553	-0.6538i$ \\
      \hline
	$2$ 	& $1$  & $2.6280	-2.0151i$   & $ - $ \\
     \hline
	$3$ 	& $0$  & $3.5772	-0.6435i$  & $3.5772	-0.6435i $ \\
      \hline
	$3$ 	& $1$   &  $3.2868	-1.9856i$     & $ - $ \\
     \hline
	$3$ 	& $2$   & $2.6940	-3.4636i$   & $ - $ \\
      \hline
	$4$ 	&  $0$  &  $4.1926	-0.6392i$     & $ 4.1926	-0.6392i $ \\
      \hline
	$4$ 	& $1$   & $3.9478	-1.9518i$   & $ - $ \\
      \hline
	$4$ 	& $2$   &  $3.4387	-3.3696i$   & $ - $ \\
     \hline
	$4$ 	& $3$   &  $2.6259	-4.9856i$  & $ - $ \\
\hline
	$5$ 	& $0$   &  $4.8034	-0.6379i$    & $ 4.8034	-0.6379i $ \\
\hline
	$5$ 	& $1$   &  $4.5933	-1.9350i$   & $ - $ \\
\hline
	$5$ 	& $2$   &  $4.1539	-3.3034i$    & $ - $ \\
\hline
	$5$ 	& $3$   &  $3.4546	-4.8171i$   & $ - $ \\
      \hline
	$5$ 	& $4$   &  $2.4810	-6.5801i$    & $ - $ \\
      \hline
      \hline
   \end{tabular}\label{fdirac15}}
   \caption{\it Massless Dirac quasinormal frequencies \(\omega r_{H}\) for $\ell=0$ to $\ell=5$.
The charges are chosen as $Q1=0$, $Q2=0.75$ and the spacetime dimension is $D=9$.}
\end{center}
      \end{table}

\newpage
\begin{table}[htb!]
\begin{center}
  {\begin{tabular}{|c|c|c|c|}
      \hline
      \hline
       $\ell$ 	& $n$ &  $6^{th}$ order WKB  & Prony  \\
      \hline
	$0$ 	& $0$  & $2.1317 -0.5964i$    &  $ 2.1317 -0.5961i $ \\
      \hline
	$1$ 	& $0$ &  $2.5279 -0.7275i$      & $ 2.5279 -0.7275i $ \\
      \hline
	$2$ 	& $0$ &   $3.1238 -0.7074i$      &  $ 3.1238 -0.7074i$ \\
      \hline
	$2$ 	& $1$  &  $2.6867 -2.1949i$   & $ - $ \\
     \hline
	$3$ 	& $0$  &  $3.7479 -0.6832i$  & $ 3.7479 -0.6832i $ \\
      \hline
	$3$ 	& $1$   &  $3.3458 -2.1324i$     & $ - $ \\
     \hline
	$3$ 	& $2$   & $2.5317 -3.6708i$   & $ - $ \\
      \hline
	$4$ 	&  $0$  & $4.3556 -0.6739i$     & $4.3556 -0.6739i $ \\
      \hline
	$4$ 	& $1$   & $4.0234 -2.0634i$   & $ - $ \\
      \hline
	$4$ 	& $2$   &  $3.3140 -3.5574i$   & $ - $ \\
     \hline
	$4$ 	& $3$   &  $2.1248 -5.1437i$  & $ - $ \\
\hline
	$5$ 	& $0$   &  $4.9538 -0.6714i$    & $ 4.9538 -0.6714i$ \\
\hline
	$5$ 	& $1$   &  $4.6733 -2.0331i$   & $ - $ \\
\hline
	$5$ 	& $2$   &  $4.0650 -3.4558i$    & $ - $ \\
\hline
	$5$ 	& $3$   &  $3.0514 -4.9886i$   & $ - $ \\
      \hline
	$5$ 	& $4$   &  $1.5429 -6.6723i$    & $ - $ \\
      \hline
      \hline
   \end{tabular}\label{fdirac16}}
   \caption{\it Massless Dirac quasinormal frequencies \(\omega r_{H}\) for $\ell=0$ to $\ell=5$.
The charges are chosen as $Q1=Q2=1$ and the spacetime dimension is $D=9$.}
\end{center}
      \end{table}

\newpage
\begin{table}[htb!]
\begin{center}
  {\begin{tabular}{|c|c|c|c|}
      \hline
      \hline
       $\ell$ 	& $n$ &  $6^{th}$ order WKB  & Prony  \\
      \hline
	$0$ 	& $0$  & $2.3363 -0.5897i$    &  $2.3364 -0.5897i$ \\
      \hline
	$1$ 	& $0$ &  $2.7332 -0.7603i$      & $2.7333 -0.7603i$ \\
      \hline
	$2$ 	& $0$ &   $3.3436 -0.7571i$      &  $ 3.3436 -0.7571i $ \\
      \hline
	$2$ 	& $1$  & $2.9325 -2.3281i$   & $ - $ \\
     \hline
	$3$ 	& $0$  & $4.0002 -0.7351i$  & $ 4.0002 -0.7351i $ \\
      \hline
	$3$ 	& $1$   &  $3.6023 -2.2977i$     & $ - $ \\
     \hline
	$3$ 	& $2$   & $2.8132 -3.9808i$   & $ - $ \\
      \hline
	$4$ 	&  $0$  &  $4.6465 -0.7245i$     & $ 4.6465 -0.7245i$ \\
      \hline
	$4$ 	& $1$   & $4.3090 -2.2305i$   & $ - $ \\
      \hline
	$4$ 	& $2$   &  $3.6093 -3.8796i$   & $ - $ \\
     \hline
	$4$ 	& $3$   &  $2.4595 -5.6953i$  & $ - $ \\
\hline
	$5$ 	& $0$   &  $5.2838 -0.7208i$    & $5.2838 -0.7208i $ \\
\hline
	$5$ 	& $1$   &  $4.9972, -2.1939i$   & $ - $ \\
\hline
	$5$ 	& $2$   &  $4.3873 -3.7659i$    & $ - $ \\
\hline
	$5$ 	& $3$   &  $3.3968 -5.5096i$   & $ - $ \\
      \hline
	$5$ 	& $4$   &  $1.9660-7.4862i$    & $ - $ \\
      \hline
      \hline
   \end{tabular}\label{fdirac17}}
   \caption{\it Massless Dirac
quasinormal frequencies \(\omega r_{H}\) for $\ell=0$ to $\ell=5$.
The charges are chosen as $Q1=0$, $Q2=1$ and the spacetime dimension is $D=9$. }
\end{center}
      \end{table}

\ \ \ \ \ \ \ \ \ \ \ \ \ \ \ \ \ \ \ \ \ \ \ \ \ \ \ \ \ \ \ \ \ \
\ \ \ \ \ \ \ \ \ \ \ \ \ \ \ \ \ \ \ \ \ \ \ \ \ \ \ \ \ \ \ \ \ \
\ \ \ \ \ \ \ \ \ \ \ \ \ \ \ \ \ \ \ \ \ \ \ \ \ \ \ \ \ \ \ \ \ \
\ \ \ \ \ \ \ \ \ \ \ \ \ \ \ \ \ \ \ \ \ \ \ \ \ \ \ \ \ \ \ \ \ \
\ \ \ \ \ \ \ \ \ \ \ \ \ \ \ \ \ \ \ \ \ \ \ \ \ \ \ \ \ \ \ \ \ \
\ \ \ \ \ \ \ \ \ \ \ \ \ \ \ \ \ \ \ \ \ \ \ \ \ \ \ \ \ \ \ \ \ \
\ \ \ \ \ \ \ \ \ \ \ \ \ \ \ \ \ \ \ \ \ \ \ \ \ \ \ \ \ \ \ \ \ \
\ \ \ \ \ \ \ \ \ \ \ \ \ \ \ \ \ \ \ \ \ \ \ \ \ \ \ \ \ \ \ \ \ \
\ \ \ \ \ \ \ \ \ \ \ \

\end{document}